\documentclass[apj]{emulateapj}
\usepackage{multirow}
\usepackage{graphics}
\usepackage{natbib}
\usepackage{footnote}
\usepackage{amsmath,amsthm}
\usepackage{longtable}
\usepackage[breaklinks,colorlinks,urlcolor=blue,citecolor=blue,linkcolor=blue]{hyperref} 
\usepackage{ulem}

\shorttitle{Unidentified type 1 AGNs}
\shortauthors{Oh et al.}	

\begin{document}

\title{A New Catalogue of Type 1 AGN and its implication on the AGN unified model}

\author{Kyuseok Oh\altaffilmark{1}, Sukyoung K. Yi\altaffilmark{2,3}, Kevin Schawinski\altaffilmark{1}, Michael Koss\altaffilmark{1,$\dagger$}, Benny Trakhtenbrot\altaffilmark{1,$\ddagger$}}
\altaffiltext{1}{Institute for Astronomy, Department of Physics, ETH Zurich, Wolfgang-Pauli-Strasse 27, CH-8093 Zurich, Switzerland; kyuseok.oh@phys.ethz.ch}
\altaffiltext{2}{Department of Astronomy, Yonsei University, Seoul 120-749, Republic of Korea} 
\altaffiltext{3}{Yonsei University Observatory, Yonsei University, Seoul 120-749, Republic of Korea; yi@yonsei.ac.kr}
\altaffiltext{$\dagger$}{\textrm{Ambizione fellow}}
\altaffiltext{$\ddagger$}{\textrm{Zwicky fellow}}

\def\OI{[\mbox{O\,{\sc i}}]~$\lambda 6300$}
\def\OIII{[\mbox{O\,{\sc iii}}]~$\lambda 5007$}
\def\OIIIs{[\mbox{O\,{\sc iii}}]~$\lambda 4363$}
\def\OIIIab{[\mbox{O\,{\sc iii}}]$\lambda\lambda 4959,5007$}
\def\SIIab{[\mbox{S\,{\sc ii}}]~$\lambda\lambda 6717,6731$}
\def\SII{[\mbox{S\,{\sc ii}}]~$\lambda \lambda 6717,6731$}
\def\NII{[\mbox{N\,{\sc ii}}]~$\lambda 6584$}
\def\NIIb{[\mbox{N\,{\sc ii}}]~$\lambda 6584$}
\def\NIIa{[\mbox{N\,{\sc ii}}]~$\lambda 6548$}
\def\NI{[\mbox{N\,{\sc i}}]~$\lambda \lambda 5198,5200$}

\def\OIIa{[\mbox{O{\sc ii}}]~$\lambda 3726$}
\def\OIIb{[\mbox{O{\sc ii}}]~$\lambda 3729$}
\def\NeIIIa{[\mbox{Ne{\sc iii}}]~$\lambda 3869$}
\def\NeIIIb{[\mbox{Ne{\sc iii}}]~$\lambda 3967$}
\def\OIIIa{[\mbox{O{\sc iii}}]~$\lambda 4959$}
\def\OIIIb{[\mbox{O{\sc iii}}]~$\lambda 5007$}
\def\HeII{{He{\sc ii}}~$\lambda 4686$}
\def\ArIVa{[\mbox{Ar{\sc iv}}]~$\lambda 4711$}
\def\ArIVb{[\mbox{Ar{\sc iv}}]~$\lambda 4740$}
\def\NIa{[\mbox{N{\sc i}}]~$\lambda 5198$}
\def\NIb{[\mbox{N{\sc i}}]~$\lambda 5200$}
\def\HeI{{He{\sc i}}~$\lambda 5876$}
\def\OI{[\mbox{O{\sc i}}]~$\lambda 6300$}
\def\OIb{[\mbox{O{\sc i}}]~$\lambda 6364$}
\def\SIIa{[\mbox{S{\sc ii}}]~$\lambda 6716$}
\def\SIIb{[\mbox{S{\sc ii}}]~$\lambda 6731$}
\def\ArIII{[\mbox{Ar{\sc iii}}]~$\lambda 7136$}

\def\Ha{{H$\alpha$}}
\def\Hb{{H$\beta$}}
\def\Hg{{H$\gamma$}}
\def\Hd{{H$\delta$}}

\def\NIIHa{[\mbox{N\,{\sc ii}}]/H$\alpha$}
\def\SIIHa{[\mbox{S\,{\sc ii}}]/H$\alpha$}
\def\OIHa{[\mbox{O\,{\sc i}}]/H$\alpha$}
\def\OIIIHb{[\mbox{O\,{\sc iii}}]/H$\beta$}

\def\NH{$N_{\textrm{H}}$}
\def\Ebmv{E($B-V$)}
\def\LOIII{$L[\mbox{O\,{\sc iii}}]$}
\def\Ledd{${L/L_{\rm Edd}}$}
\def\LOIIIs4{$L[\mbox{O\,{\sc iii}}]$/$\sigma^4$}
\def\LOIIIMbh{$L[\mbox{O\,{\sc iii}}]$/$M_{\rm BH}$}
\def\Mbh{$M_{\rm BH}$}
\def\Msigma{$M_{\rm BH} - \sigma$}
\def\Ms{$M_{\rm *}$}
\def\Msun{$M_{\odot}$}
\def\Msunyr{${\rm M_{\odot}yr^{-1}}$}

\def\ergs{${\rm erg}~{\rm s}^{-1}$}
\def\kms{${\rm km}~{\rm s}^{-1}$}
\newcommand{\cms}{\mbox{${\rm cm\;s^{-1}}$}}
\newcommand{\pccm}{\mbox{${\rm cm^{-3}}$}}
\newcommand{\sauron}{{\texttt {SAURON}}}
\newcommand{\oasis}{{\texttt {OASIS}}}
\newcommand{\HST}{{\it HST\/}}

\newcommand{\Vg}{$V_{\rm gas}$}
\newcommand{\Sg}{$\sigma_{\rm gas}$}
\newcommand{\eg}{e.g.,}
\newcommand{\ie}{i.e.,}

\newcommand{\gandalf}{{\texttt {gandalf}}}
\newcommand{\fracDeV}{{\texttt {FracDeV}}} 
\newcommand{\ppxf}{{\texttt {pPXF}}}

\begin{abstract}

We have newly identified a substantial number of type 1 active galactic nuclei (AGN) featuring weak broad-line regions (BLRs) at $z<0.2$ from detailed analysis of galaxy spectra in the Sloan Digital Sky Survey Data Release 7. These objects predominantly show a stellar continuum but also a broad \Ha\ emission line, indicating the presence of a low-luminosity AGN oriented so that we are viewing the central engine directly without significant obscuration. These accreting black holes have previously eluded detection due to their weak nature. The new BLR AGNs we found increased the number of known type 1 AGNs by 49\%. Some of these new BLR AGNs were detected at the \textit{Chandra} X-ray Observatory, and their X-ray properties confirm that they are indeed type 1 AGN. Based on our new and more complete catalogue of type 1 AGNs, we derived the type 1 fraction of AGNs as a function of \OIII\ emission luminosity and explored the possible dilution effect on the obscured AGN due to star-formation. The new type 1 AGN fraction shows much more complex behavior with respect to black hole mass and bolometric luminosity than suggested by the existing receding torus model. The type 1 AGN fraction is sensitive to both of these factors, and there seems to be a sweet spot (ridge) in the diagram of black hole mass and bolometric luminosity. Furthermore, we present a hint that the Eddington ratio plays a role in determining the opening angles. 

\end{abstract}	


\section{Introduction}

According to the unification scheme of active galactic nuclei (AGNs, \citealt{ant93, urr95}), which is widely accepted in current astrophysics, different viewing angles explain the variety of phenomenological subclasses of AGNs. For example, the optical spectra of type 1 AGNs present broad permitted emission lines as well as narrow emission lines exposing their central nuclear region, whereas type 2 AGNs show only narrow emission lines. According to the simplest AGN unification model and its strictest interpretation, the feasibility of observing broad permitted emission lines depends solely on the angle between the observer's line of sight and the axis of an obscuring medium, often called the ``dust torus''. 

With the aid of mega-scale surveys such as the Sloan Digital Sky Survey (SDSS; \citealt{yor00}) that provide homogenous photometric and spectroscopic data, AGNs have become one of the most actively studied topics in astrophysics. In particular, studies on {\it obscured AGNs} showing narrow emission lines have shed light on the connection between the host galaxies and the central AGN \citep{kau03, sal07, sch07, wes07, sch09, hag10, scha10}.

{\it Unobscured} (type 1) AGNs reveal their bright nuclear light directly, and it has been an important task and subject of interest to decompose the observed light into a central nuclear component and a diffuse galaxy component. For example, the high-spatial-resolution Advanced Camera for Surveys on the \textit{Hubble Space Telescope}, has been extensively used for this purpose \citep{san04,  bal07, alo08, sim08, sch11, sim13} in combination with  two-dimensional surface brightness fitting techniques \citep{pen02}.

\begin{figure*}
\centering							      
\includegraphics[width=1.0\textwidth]{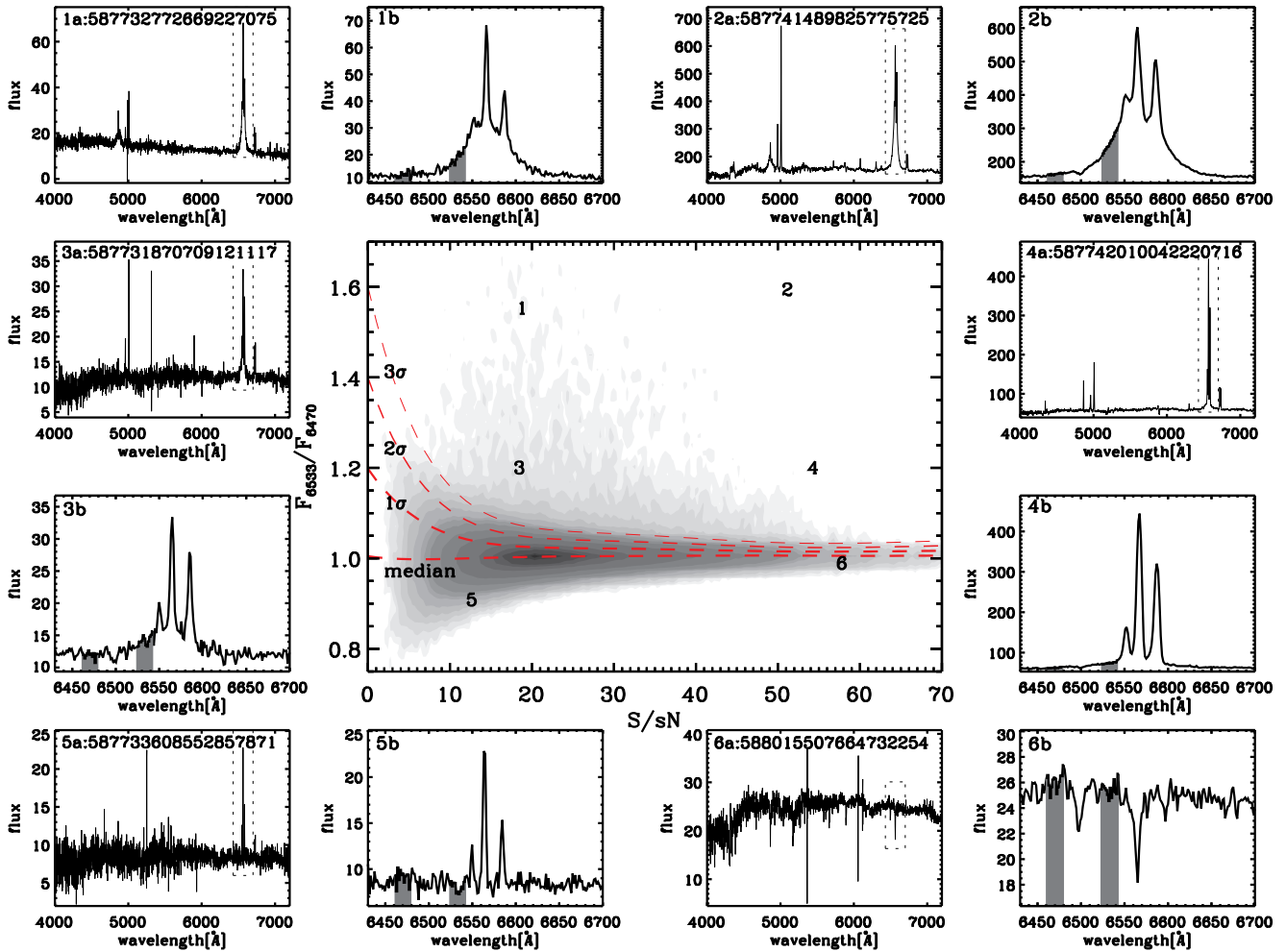} 
\caption{Flux ratio ($\rm{F_{6,533}/F_{6,470}}$) distribution versus signal-to-statistical noise (S/sN). Demarcation lines in the central panel are shown as red dashed lines. The three lines indicate 84.1\%, 97.7\%, and 99.8\% distributions. Example spectra represented by the numbered markings in the central panel are shown in the surrounding panels, with two small sub panels for each object. Surrounding panels marked with `a' on the top left side show the overall shape of the spectrum, whereas the panels marked `b' show the \Ha\ region. Each object is labeled in sub panel `a' by SDSS ObjID. Black solid lines indicate the observed spectrum. The two narrow bands that define the flux ratio are shaded gray. Four spectra (numbered from 1 to 4) are chosen as type 1 AGN candidates out of six examples by the demarcation lines. Flux has the unit [10$^{-17} \rm{erg s^{-1} cm^{-2}\AA^{-1}}$]. 
}
\label{flux_ratio}
\end{figure*}

The fraction of type 1 AGN (among all AGNs) can act as an effective test of any model for AGN classifications and can be used to extract and constrain critical information, such as the opening angle and source luminosity, in the assumed framework. Based on a sample of high-luminosity and radio-selected broad-line AGNs, \citet{wil00} found the type 1 AGN fraction of 40\% and a typical value of the ``half-opening'' angle of about $53^{\circ}$. The precise value of the type 1 AGN fraction has been reported to be substantially lower than this finding, when low-redshift surveys of lower optical luminosity cuts are used instead \citep{ost88, huc92, mai95, mai03}. This difference seems to be sensible, as more recent, multiwavelength studies found a strong, positive luminosity dependence of the type 1 AGN fraction \citep{ste03, hao05, sim05, has08, bur11, ass13}.
 
AGNs exhibit characteristic spectral features across the electromagnetic spectrum: e.g., \OIII\ optical emission lines, mid-infrared and hard X-ray luminosity as isotropic quantities tracing the source luminosity, which are naturally employed by many type 1 AGN fraction studies. In particular, \citet{sim05} found that the type 1 AGN fraction gradually increases with \LOIII\ based on SDSS Data Release 2 \citep{aba04} and hard X-ray data in the literature \citep{ued03, gri04, has04}. 

A theoretical framework has been claimed to explain such a luminosity dependency of the type 1 AGN fraction. For example, the receding torus model \citep{law91, fal95, hil96, sim05} imposes a dust sublimation radius of the torus, which is directly affected by the nuclear luminosity. As the luminous central engine pushes the dust torus away, thereby increasing the torus' inner radius, the opening angle must increase if the height of the torus is fixed. This naturally explains the observed luminosity dependence of the type 1 AGN fraction.  

A precise determination of the type 1 AGN fraction is obviously a key test of a model's validity. Since the AGN unified model explains the observability of both types of AGN as the consequences of obscuration induced by the geometry and the structure of AGN, the frequency of observed type 1 AGN indirectly constrains the model. For instance, higher type 1 AGN fraction implies wider opening angle, smaller scale height of the torus, and higher nuclear luminosity. Herein we report on our efforts to discover new type 1 AGNs from the SDSS database to provide such a test. We describe how we found a substantial number of new type 1 AGNs and derived a more up-to-date type 1 AGN fraction. The use of the new type 1 AGN fraction places a harsh constraint on the model, and if the scheme used herein proves convincing, it appears that an important modification to the model will be necessary. We release a new catalogue of type 1 AGNs in the nearby universe by means of this paper.

The paper is organized as follows. In Section 2, we describe the selection technique whereby we constructed the new type 1 AGN database, and describe a simulation that quantified the detection efficiency of our method. We demonstrate our conservative scheme for type 1 selection and determine the final type 1 AGN sources, taking the simulation into account. In Section 3, we present the new type 1 AGN fraction and discuss how contamination caused by star formation activities may affect it. Finally, we discuss our findings and summarize our results in Section 4. 

We assume a cosmology with $h$ = 0.70, $\Omega_{\rm{M}}$ = 0.30, and $\Omega_{\rm{\Lambda}}$ = 0.70 throughout this work.

\begin{figure*}
\centering							
\includegraphics[width=1.0\textwidth]{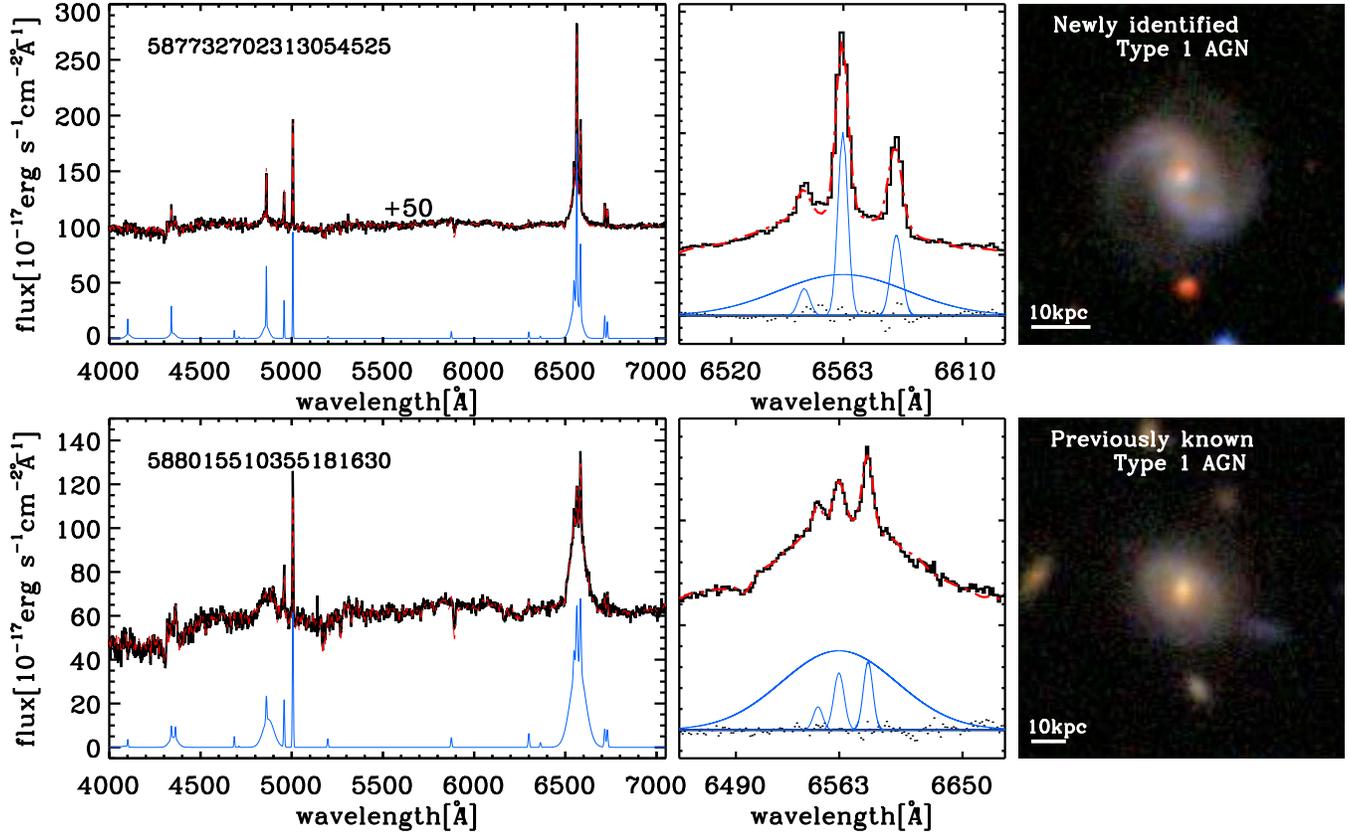} 
\caption{Example spectra with SDSS \textit{gri} composite images for a newly identified type 1 AGN (top) 
and a previously known type 1 AGN from the SDSS pipeline (bottom). 
Left and middle panels present the overall spectra and their \Ha\ regions, respectively. 
Black lines indicate the observed spectra and red dashed lines include stellar continuum fits. 
Note that the observed spectrum and the fit in the top panel are shown with offset in the ordinate for clarity (+50 in flux unit).
Detected emission lines are shown in the left panels as blue lines. 
In particular, broad \Ha\ components are shown in the middle panels as well as other narrow emission lines. 
Dots in the middle panels show the residuals. 
}
\label{SED_example_fig2}
\end{figure*}

\section{Type I AGN Sample Selection}
\label{sec:data}

\subsection{Selection based on broad \Ha\ features}
\label{ssec:sample}

The presence of unidentified type 1 AGNs in the nearby universe has been reported by \citet{hao05}, \citet{gre07}, and \citet{ oh11} [a.k.a. the OSSY catalogue]. In particular, \citet{oh11} found that the presence of broad Balmer emission lines were responsible for a very poor fit by model to both the emission lines and the stellar continuum during their construction of the OSSY catalogue. The authors used a flux ratio near the \Ha\ emission line to systematically search for BLR AGNs from the SDSS DR7 galaxy database. In this work, we adopted their basic scheme but developed it further to minimize dubious detections. 

The OSSY catalogue provides new spectral line strength measurements for the SDSS DR7 galaxies at $z<0.2$ (N = 664,187). By combining stellar kinematics measurements (\ppxf; \citealt{cap04}) and spectral line fitting algorithm (\gandalf; \citealt{sar06}), the authors released improved stellar absorption and emission-line strengths, stellar velocity dispersion, and quality assessing parameters. We start our analysis based on the OSSY catalogue. 

To identify the spectra showing broad \Ha\ features\footnote{broad \Ha\ is also a signature of supernovae but it is known as rare \citep{fil97}.}, we computed a ratio between the mean fluxes at two specific wavelength bands. The first narrow band (6,460 $\--$ 6,480 \AA) is located near \Ha\ but far enough from it to be free from the \Ha\ broad emission, mainly to secure the pseudo-continuum for \Ha. The second (6,523 $\--$ 6,543 \AA) was placed close to \NIIa\ to highlight the broad \Ha\ feature. Then, we computed a flux ratio ($\rm{F_{6,533}/F_{6,470}}$) based on the two bands. Fig.~\ref{flux_ratio} shows the flux ratio versus the signal-to-statistical noise ratio (S/sN) of the chosen specific wavelength bands for all galaxies listed in the OSSY catalogue (N = 664,187; $z<0.2$). We used this diagram to first detect \Ha\ BLR candidates. Examples of spectral energy distributions (SEDs) for various values of flux ratios are illustrated in Fig.~\ref{flux_ratio} (surrounding panels), including the overall spectra and the \Ha\ regions. As the flux ratio increases, broad \Ha\ features become more prominent (see objects marked 1, 3, and 5 in order of decreasing broad \Ha\ strength), and the same trend is visible in the higher S/sN regime (objects 2, 4, and 6). It seems clear that the flux ratio is effective for detecting \Ha\ BLR candidates. 
\begin{figure*}
\centering
\includegraphics[width=0.9\textwidth]{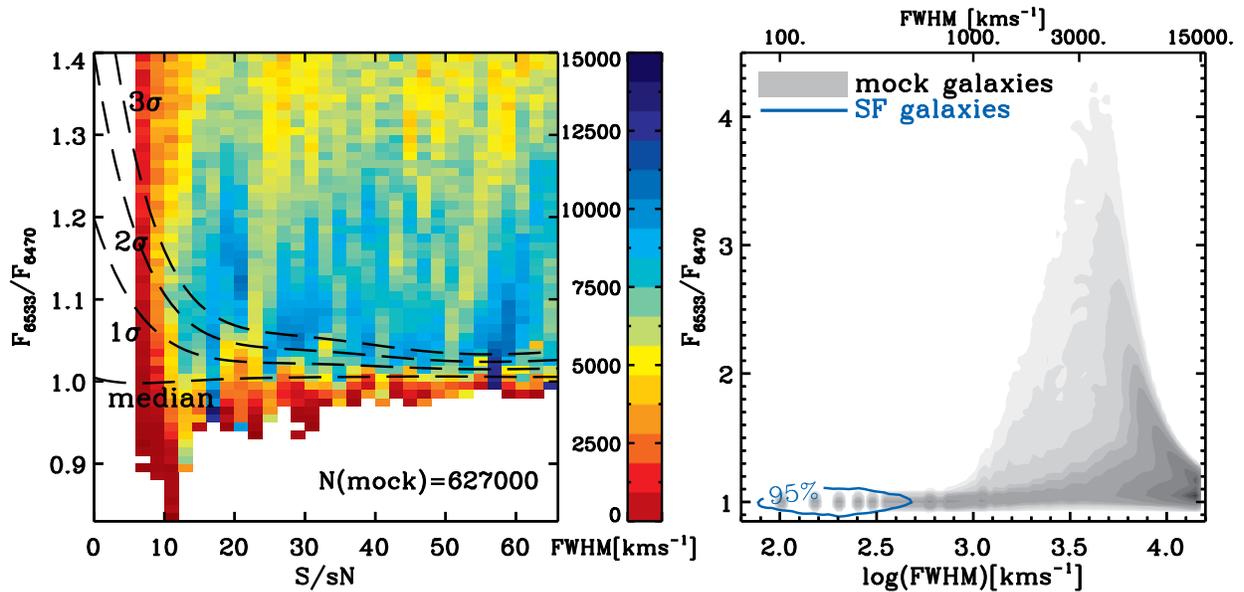} 
\caption{Flux ratio distribution with respect to S/sN (left) and FWHM (right) for the simulated galaxies. 
In the left panel, color grids and the corresponding color bar represent the mean FWHM. 
In the right panel, FWHMs measured for the actively star-forming galaxies are indicated by a blue contour (95$\%$) while the values for the mock spectra are indicated by achromatic filled contours.
}
\label{simul_results}
\end{figure*}

The 1$\sigma$ demarcation line shown in Fig.~\ref{flux_ratio} was applied to first select type 1 AGN candidates. A total of 17\% of galaxies have a flux ratio above this cut. We used \gandalf(Gas and Absorption Line Fitting, \citealt{sar06}) to fit the spectrum based on both nebular emissions and stellar absorptions. Our approach for analyzing the SDSS spectra can be summarized as a two-step process. First, we masked nebular emission lines and derived the stellar kinematics by directly matching the observed spectrum with stellar templates by using the \ppxf\ IDL program \citep{cap04}. For the ``stellar'' templates, we used the \citet{bru03} stellar population synthesis models and the MILES stellar library \citep{san06}. After measuring the kinematic broadening, we lifted the masking on the nebular emission lines and simultaneously fit stellar continuum and emission lines by using the Levenberg-Marquardt minimization \citep{mar09}\footnote{\ppxf\ and \gandalf\ use the MPFIT IDL routine provided by Craig Markwardt at http://cow.physics.wisc.edu/\~{}craigm/idl/}. 

The standard \ppxf\ and \gandalf\ procedures generally work well on normal galaxies, but a more elaborate consideration is necessary for type 1 AGNs. For example only for type 1 AGNs, we adopted a wide masking width of 12,000 \kms\ for broad Balmer lines (\Ha, \Hb, \Hg, and \Hd) and 1,200 \kms\ for narrow emission lines, prior to measuring stellar kinematics. Besides, we imposed an initial guess of a large width, 1,000 \kms, for broad Balmer component fitting. We implemented single broad Gaussian component for each Balmer line. As demonstrated in the middle panels of Fig.~\ref{SED_example_fig2}, these prescriptions were effective for decomposing the broad (\Ha) and narrow (\Ha, \NIIa\ and \NIIb) components around the broad \Ha\ region. 

\begin{center}
\begin{deluxetable}{rlcc}
\tabletypesize{\scriptsize}
\tablecaption{Simulation designs}
\tablewidth{0pt}
\tablehead{
\colhead{components} &
\colhead{range} &
\colhead{increment} &
\colhead{No. of bins} 
}
\startdata
S/sN\tablenotemark{a}				& $5 \-- 55, > 55$ 													& 5 		& 11 	\\ 
FWHM 							& $50 \-- 15,000$ [\kms]												& 50 	& 300  	\\
\multirow{2}{*}{Peak\tablenotemark{b}} 	& $0.01\times A_{{\rm H\alpha}}\tablenotemark{c} \-- 0.09\times A_{{\rm H\alpha}}$ 	& 0.01 	& 9    	\\
								& $0.1\times A_{{\rm H\alpha}} \-- 1.0\times A_{{\rm H\alpha}}$ 					&  0.1 	& 10 
\enddata
\tablecomments{In total, 627,000 mock galaxies were generated.}
\label{tab:simulation_design}
\tablenotetext{a}{Average S/sN from 4 continuum bands (4,500\--4,700, 5,400\--5,600, 6,000\--6,200, 6,800\--7,000 \AA)}
\tablenotetext{b}{Peak of the Gaussian}
\tablenotetext{c}{Amplitude of the narrow \Ha}
\end{deluxetable}
\end{center}

Following these routines, we measured the full-width at half-maximum (FWHM), flux, and amplitude over noise (A/N) ratio\footnote{A/N is defined by the ratio between Gaussian amplitude of the emission line and dispersion of continuum residual of the given spectrum} of all lines listed in the OSSY catalogue. Our selection criteria for type 1 AGN candidates are as follows. 
\begin{itemize}
	\item $0.00<z<0.20$
	\item FWHM of broad \Ha\ $>$ 800 \kms
	\item A/N of broad \Ha\ $>$ 3
\end{itemize}
The redshift limit is constrained by the wavelength of the red end of the pseudo-continuum around \Ha. We chose a cut for FWHM(\Ha) that was slightly lower than the one more widely used, 1,000 \kms\ \citep{van06, sch10, stern12}, hoping to detect even weaker BLR AGNs. As a result, we found 9,671 (8.6\%) type 1 AGN candidates among the 111,824 sources that satisfied the 1$\sigma$ cut shown in the central panel of Fig.~\ref{flux_ratio}. Of these 9,671, only 120 were of FWHM(broad \Ha) $<$ 1,000 \kms, and hence the choice of the 800 \kms\ cutoff instead of the more typical 1,000 \kms\ cutoff influenced the results and conclusions little.

\begin{center}
\begin{deluxetable}{ccccc}
\tabletypesize{\scriptsize}
\tablecaption{Detection efficiency for each $\sigma$ cut}
\tablewidth{0pt}
\tablehead{
\colhead{\multirow{2}{*}{$\sigma$ cut\tablenotemark{a}}} &
\colhead{No. of} &
\colhead{No. of} &
\colhead{completeness\tablenotemark{b}} &
\colhead{purity} \\
\colhead{} &
\colhead{mock galaxies} &
\colhead{mock BLRs} &
\colhead{(\%)} &
\colhead{(\%)} 
}
\startdata
$>1\sigma$	& 547,495	& 543,052	& 91.49	& 99.19\\ 
$>2\sigma$	& 515,939	& 514,562	& 86.69	& 99.73\\
$>3\sigma$	& 487,943	& 487,512	& 82.13	& 99.91
\enddata
\label{tab:simul_result}
\tablenotetext{a}{$\sigma$ cut marked in the central panel of Fig.~\ref{flux_ratio}}
\tablenotetext{b}{Number of all mock broad-line AGNs ($\rm{FWHM>800}$ \kms) = 593,560}
\end{deluxetable}
\end{center}
\begin{figure*}
\centering
\includegraphics[width=0.8\textwidth]{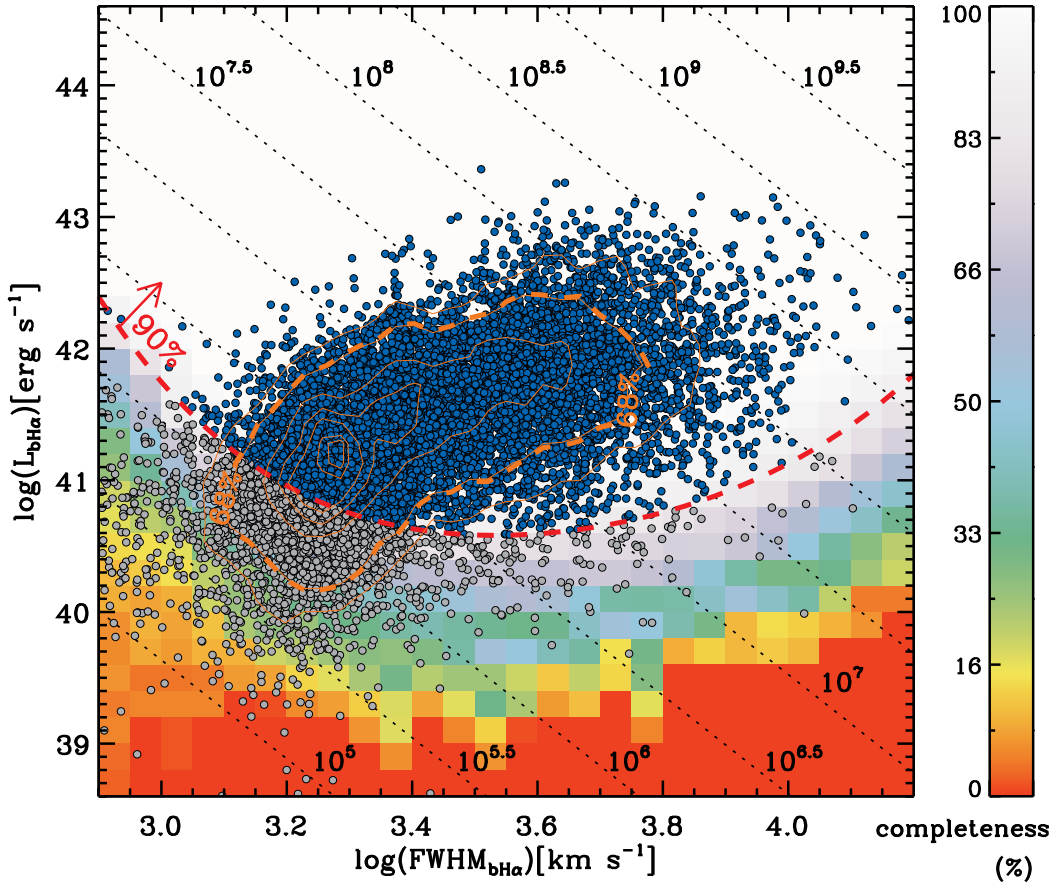} 
\caption{Luminosity versus FWHM for broad \Ha\ components. 
Completeness derived from a simulation is indicated by the underlying color grid. 
Labeled black dotted lines represent \Mbh\ in units of solar mass following the formulae developed by \citet{gre05}.
The red dashed curve indicates the demarcation used to choose type 1 AGN sources based on a 90\%\ completeness criterion.
The type 1 AGNs (N=7,619) selected by the 90\%\ completeness cut are shown as blue-filled dots; excluded points (N=2,052) below the curve are shown as gray-filled dots. 
}
\label{lum_fwhm_Ha}
\end{figure*}

\subsection{Test of the selection scheme}
\label{ssec:simulation}

The selection process described above relies on the specific flux ratio as a proxy to the broad \Ha\ feature. To verify our approach and measure the detection efficiency, that is, its completeness and purity, we conducted a simulation by generating mock spectra, each with an artificially broadened \Ha\ component.  

The basis of the mock spectra was 110 star-forming galaxies with narrow emission lines: 10 galaxies randomly chosen from each of 11 S/sN bins. To secure a wide range of the amplitude of \Ha, we used the spectra of actively star-forming galaxies of SFR $>$ 0.5 \Msunyr\ that were identified by the BPT diagnostics diagram \citep{bal81} and the Kennicutt law \citep{ken98}. We then add broad line components by varying two parameters: Gaussian peak and FWHM, while the center of the Gaussian was fixed at 6562.8 \AA. The peak of the Gaussian was fractionally scaled by the amplitude of the narrow \Ha\ emission line ($A_{\textrm{H}\alpha}$) in 19 steps, namely, by 1$\%$ increments from 1$\%$ to 10$\%$ and a 10$\%$ increment from 10$\%$ to 100$\%$ in $A_{\textrm{H}\alpha}$. The FWHM of the broad \Ha\ line was varied between 50 and 15,000 \kms\ in 50 \kms\ increments. For reference, the typical range of FWHM(\Ha) of star-forming galaxies is 50$\--$350 \kms\ within the 2$\sigma$ level. In total, 627,000 ($110 \times 19 \times 300$) mock spectra were generated, as summarized in Tab.~\ref{tab:simulation_design}.

\begin{center}
\begin{deluxetable}{ccccc}
\tabletypesize{\scriptsize}
\tablecaption{Detection efficiency for each $\sigma$ cut after applying areal ratio}
\tablewidth{0pt}
\tablehead{
\colhead{\multirow{2}{*}{$\sigma$ cut}} &
\colhead{No. of} &
\colhead{No. of} &
\colhead{completeness} &
\colhead{purity} \\
\colhead{} &
\colhead{mock galaxies} &
\colhead{mock BLRs} &
\colhead{(\%)} &
\colhead{(\%)} 
}
\startdata
$>1\sigma$	& 547,495	& 484,791	& 81.68	& 88.55\\ 
$>2\sigma$	& 515,939	& 476,341	& 80.25	& 92.33\\
$>3\sigma$	& 487,943	& 463,475	& 78.08	& 94.99
\enddata
\label{tab:simul_result_areal_ratio}
\end{deluxetable}
\end{center}
\begin{figure*}
\centering							
\includegraphics[width=1.0\textwidth]{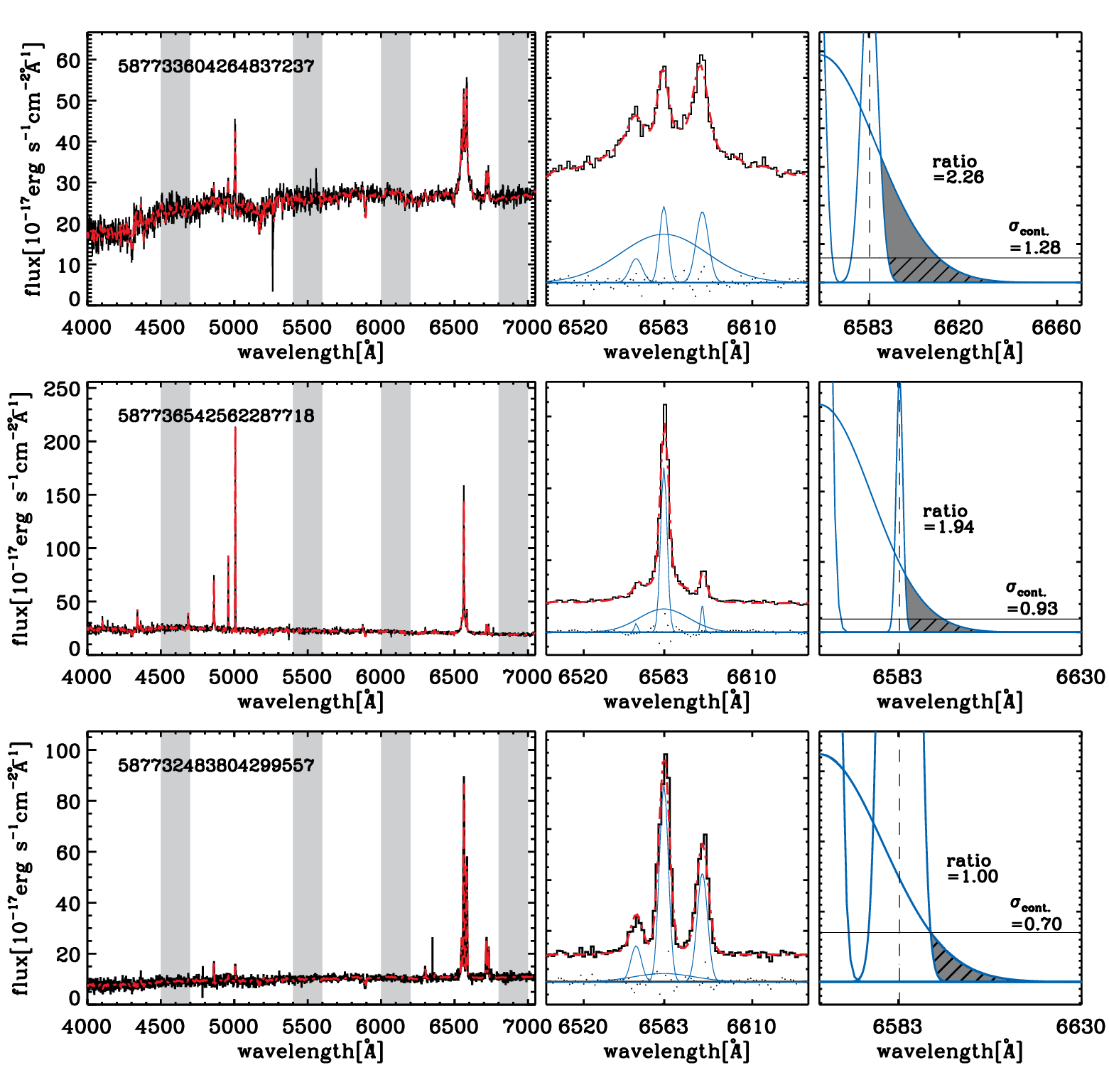} 
\caption{Example spectra illustrating the conservative selection criterion based on areal excess. 
The overall spectra are shown in the left panels, including SDSS ObjID labels at the top left. 
Black and red lines respectively indicate the observed spectrum and our fit. 
The regions that are used to measure the averaged continuum dispersion are shaded gray. 
The middle panels present the detected \NIIa, narrow \Ha, broad \Ha, and \NIIb\ emission lines (blue Gaussians), and their residuals (black dots). 
The right panels enlarge the middle panels, comparing the measured broad \Ha\ exceeding \NIIb\ (gray shades) to the level of the continuum dispersion (hatched regions with black horizontal lines). 
The areal excess, that is, the ratio between the gray shaded and hatched regions, is marked, as well as the level of the continuum dispersion, $\sigma_{cont.}$.
}
\label{SED_example}
\end{figure*}

Fig.~\ref{simul_results} shows flux ratios versus S/sN (left panel) and FWHM (right panel) for the simulated galaxies. In the left panel, the galaxies with low FWHMs (dark red color) were generally found below the 1$\sigma$ demarcation line and/or had low S/sN. As the color grids show, our 1$\sigma$ cut for recovering BLRs with a large value of FWHM (blue color) seems effective, although it still suffers from contaminants such as objects having noise fluctuations or having a broad \Ha\ component that is too shallow. We will discuss how these contaminants can be removed in the next section. The right panel of Fig.~\ref{simul_results} shows that the highly star-forming galaxies having only narrow emission lines obviously occupy the region of low flux ratio ($<$ 1.1) and low FWHM ($<$ 500 \kms). Below the FWHM of 500 \kms, the mock galaxies have almost the same flux ratios as narrow emission line galaxies, which means that it is not easy to distinguish between weak type 1 AGNs and star-forming galaxies. On the other hand, the flux ratio of the mock galaxies begins to spread out greatly for $\rm FWHM>1,000$ \kms. 

\begin{figure}
\centering							
\includegraphics[width=0.5\textwidth]{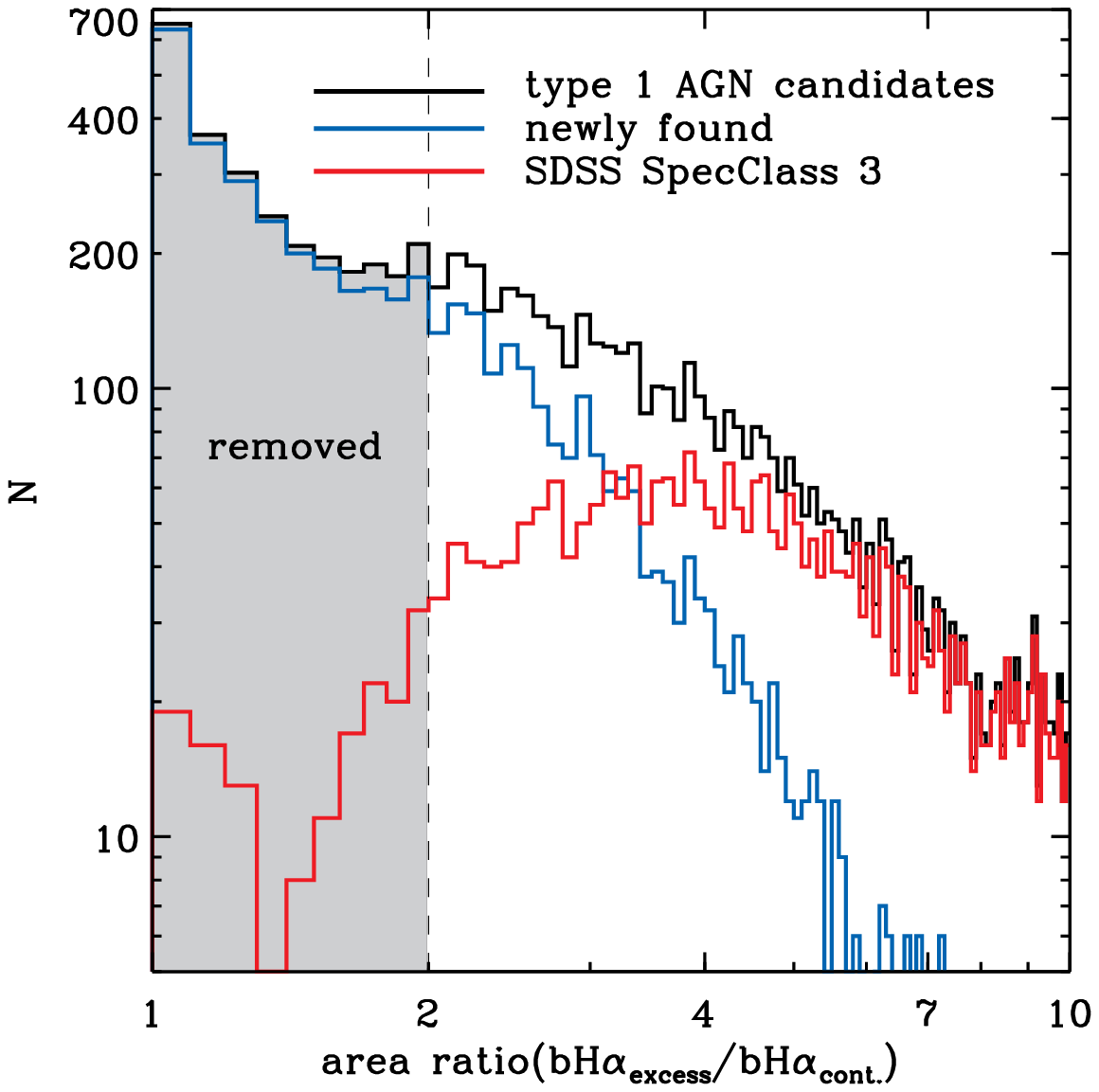} 
\caption{Histograms of area ratio for all type 1 AGN candidates (black), newly found type 1 AGNs (blue), and SDSS \textit{SpecClass} 3 sources (red). 
The new selection criterion, area ratio $>2$, is marked using a vertical dashed line; it removes 2,066 dubious candidates (gray shaded area). 
}
\label{areal_ratio_histogram}
\end{figure}

Detection efficiency (completeness and purity) can be measured by assuming that a BLR AGN has FWHM(broad \Ha) $> 800$ \kms. Completeness is defined as the fraction of mock BLR AGNs that are recovered by the procedure (i.e., the $\sigma$ cut). Purity describes the fraction of BLR AGNs among all galaxies selected by a $\sigma$ cut. Tab.~\ref{tab:simul_result} lists the completeness and purity of BLR AGNs for three choices of $\sigma$ cut. As higher $\sigma$ cuts are applied, completeness decreases but purity increases. 

Fig.~\ref{lum_fwhm_Ha} illustrates that the completeness (color key) of our selection method based on the flux ratio of the broad \Ha\ component is mainly determined by the broad \Ha\ luminosity (ordinate). We used this completeness distribution to find BLR AGN candidates most economically. We first located the locus of 90\%\ completeness pixels in the diagram. As most of the pixels above this locus satisfy the completeness of 90\%, the total completeness when using this 90\% cut is in practice much greater than 90\%. Roughly 21\% (2,052 out of 9,671) of the BLR AGN candidates (by the 1$\sigma$ criterion) were below the 90\% demarcation line and thus were missed by this scheme.

\subsection{Selection based on the broadening around ${\rm [NII]\lambda 6584}$}
\label{ssec:areal_excess}

We have so far described our selection criterion based on the broad \Ha\ feature, but there still are complexities. For instance, an extremely shallow broad \Ha\ component can be technically classified as type 1 AGN as long as it satisfies our conditions on A/N and FWHM but not without doubt. For another example, the credibility of type 1 classification for border-line objects of FWHM(\Ha) just over 800 \kms\ can be doubted, because it is difficult to measure the width of such a marginally-broad \Ha\ component accurately, especially in the midst of strong neighboring narrow emission lines (\NII). 

Having carried out a sophisticated spectral fit for each galaxy spectrum and having determined decomposed broad \Ha\ properties as a result, we were able to perform a more stringent type 1 classification than carried out so far. We compared the area of the broad \Ha\ component in question with the noise level of the observed continuum. The area of broad \Ha\ beyond \NIIb\ to longer wavelengths can be used as an alternative measure to FWHM for the broad \Ha\ feature. Meanwhile, the area defined by the dispersion (measurement uncertainty) of the continuum can be used as a control parameter. Fig.~\ref{SED_example} provides an illustration with example spectra. The right panels demonstrate how we assess the broad \Ha\ component (gray shaded area) that surpasses the level of the continuum dispersion (hatched area), which is marked as $\sigma_{cont.}$. As an additional type 1 AGN criterion, we take the areal ratio of the broad \Ha\ over the measurement uncertainty around \NIIb\ greater than 2. The application of this new cut removed 3,940 controversial type 1 AGN candidates (Fig.~\ref{areal_ratio_histogram}). Tab.~\ref{tab:simul_result_areal_ratio} lists the completeness and purity for each $\sigma$ cut after implementing areal ratio analysis. As is the case with any other criterion, some of the objects removed in this step are likely to be true type 1 AGN. Yet, we felt it necessary to apply this additional criterion to construct a more robust list of type 1 AGN candidates. 

\begin{figure}
\centering							
\includegraphics[width=0.5\textwidth]{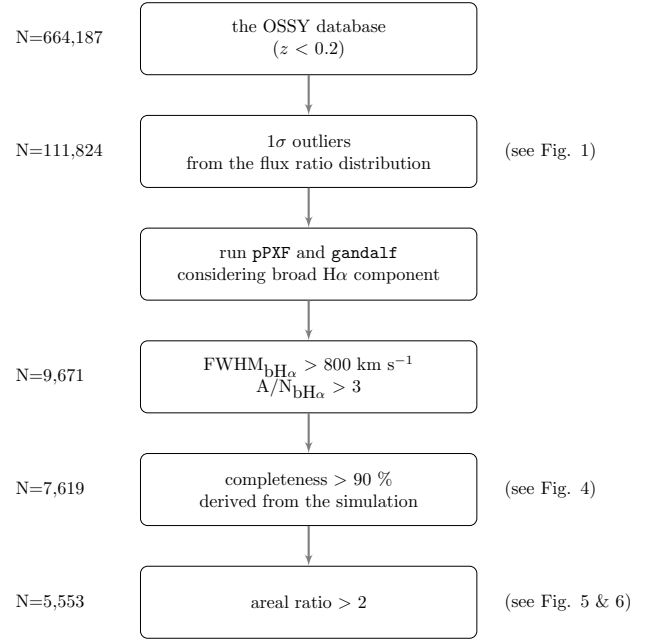} 
\caption{Flowchart of the type 1 AGN selection.
}
\label{flow_chart}
\end{figure}

By using the completeness ($>90$\%) and areal ratio ($>2$) arguments, we identified 5,553 type 1 AGN candidates. It should be noted that the majority (91.3\%) of the type 1 AGN candidates below the 90\% completeness cut in Fig.~\ref{lum_fwhm_Ha} were eliminated through the areal excess test, confirming the effectiveness of the completeness test as a filtering scheme. The flowchart in Fig.~\ref{flow_chart} shows the number of objects that survived each selection. The broad nature of the \Ha\ emission line of type 1 AGNs eluded our initial attempt to measure their strengths. Perhaps for the same reason, many (38\%, 700 out of 1,835) of our newly found type 1 AGNs are not included in the MPA-JHU catalogue\footnote{http://www.mpa-garching.mpg.de/SDSS/DR7/}. Hereinafter, our analysis is based on these 5,553 sources. 

We list the measured properties of these sources as well as the estimated quantities with designation flag in Tab.~\ref{tab:all_type1}. We used the formula developed by \citet{gre05} that depends on the line width and luminosity of the broad \Ha\ to estimate black hole mass. We assumed bolometric luminosity by adopting the relationship developed by \citet{hec04} ($L_{\textrm{bol}} \approx$ 3,500\LOIII). The cross-matched flag presents that 456 type 1 AGNs (8.2\%) and 2,954 (53.2\%) sources in our catalogue are found in \citet{hao05} and \citet{gre07}, respectively.

\begin{center}
\begin{deluxetable}{lllrr}
\tabletypesize{\scriptsize}
\tablecaption{Detection of \textit{SpecClass} 3 sources}
\tablewidth{0pt}
\tablehead{
\colhead{} &
\colhead{} &
\colhead{N} &
\colhead{\%} 
}
\startdata
\multicolumn{2}{l}{total}														& 4125 				& 100.00 \\
\multicolumn{2}{l}{detected\tablenotemark{a}}										&      				&           \\
& \multicolumn{1}{l}{after flux ratio cut} 											& 4079				& 98.89 \\
& \multicolumn{1}{l}{after A/N$_{\textrm{bH}\alpha}$ \&\ FWHM$_{\textrm{bH}\alpha}$ cut}		& 3969				& 96.22 \\
& \multicolumn{1}{l}{after completeness cut ($>$90\%)\tablenotemark{b}} 					& 3921				& 95.05 \\
& \multicolumn{1}{l}{ater areal excess cut ($>$100\%)\tablenotemark{c}}					& 3718\tablenotemark{d}	& 90.13 \\
\multicolumn{2}{l}{missed}														& 407 				& 9.87 \\
\sidehead{Description on the sources that failed flux ratio cut}
& \multicolumn{1}{l}{only narrow emission lines}										&	19	&	41.30 \\
& \multicolumn{1}{l}{absence of spectral information}									&	11	& 	23.91 \\
& \multicolumn{1}{l}{maybe broad-line AGNs}										&	10	& 	21.74 \\
& \multicolumn{1}{r}{passing A/N$_{\textrm{bH}\alpha}$ \&\ FWHM$_{\textrm{bH}\alpha}$ cut}	&	(10)	&	(21.74) \\
& \multicolumn{1}{r}{passing completeness cut ($>$90\%)}								&	(8)	&	(17.39) \\
& \multicolumn{1}{r}{passing areal excess cut ($>$100\%)}								&	(5)	&	(10.87) \\
& \multicolumn{1}{l}{asymmetric bump on \Ha}										&	3   	&	6.52	   \\
& \multicolumn{1}{l}{non-emission lines}											&	2	&	4.35   \\
& \multicolumn{1}{l}{unknown spectrum}											&	1	&	2.17 
\enddata
\label{tab:specclass}
\tablenotetext{a}{Number of all detected \textit{SpecClass} 3 sources chosen from the flux ratio analysis at $z<0.2$}
\tablenotetext{b}{see Section 2.3 and Fig.~\ref{lum_fwhm_Ha}}
\tablenotetext{c}{see Section 2.4 and Fig.~\ref{SED_example}}
\tablenotetext{d}{Number of \textit{SpecClass} 3 sources in the final catalog}
\end{deluxetable}
\end{center}
\begin{figure}
\centering							
\includegraphics[width=0.5\textwidth]{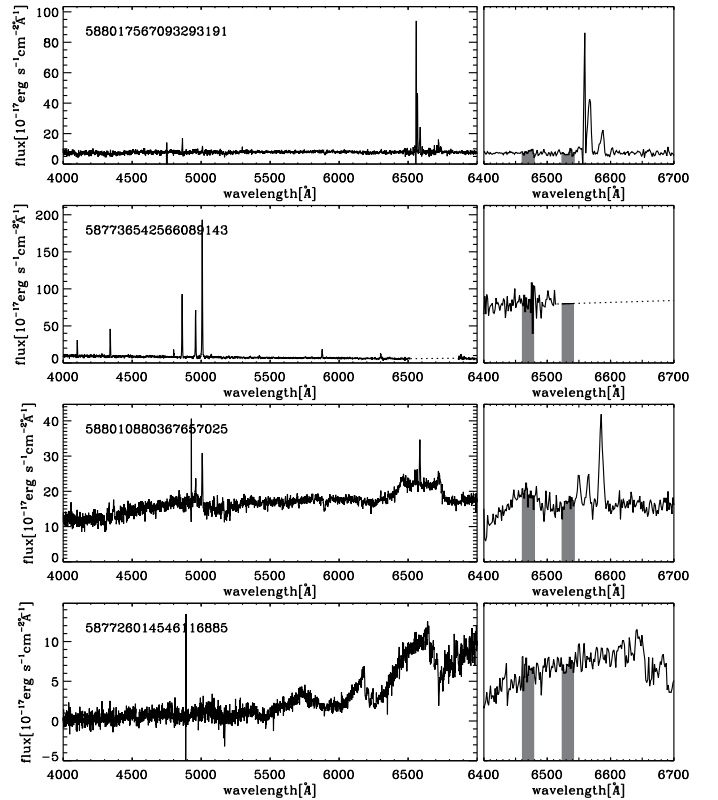} 
\caption{Spectra of example \textit{SpecClass} 3 sources below the 1$\sigma$ demarcation line, for a narrow-emission-line galaxy (top), an incomplete spectrum near \Ha\ (upper), an asymmetric bump feature (lower), and an unknown source (bottom). Right panels show the \Ha\ region; the two narrow bands that define the flux ratio are shaded gray. 
}
\label{example_spec3_sed}
\end{figure}
\begin{figure*}
\centering
\includegraphics[width=0.8\textwidth]{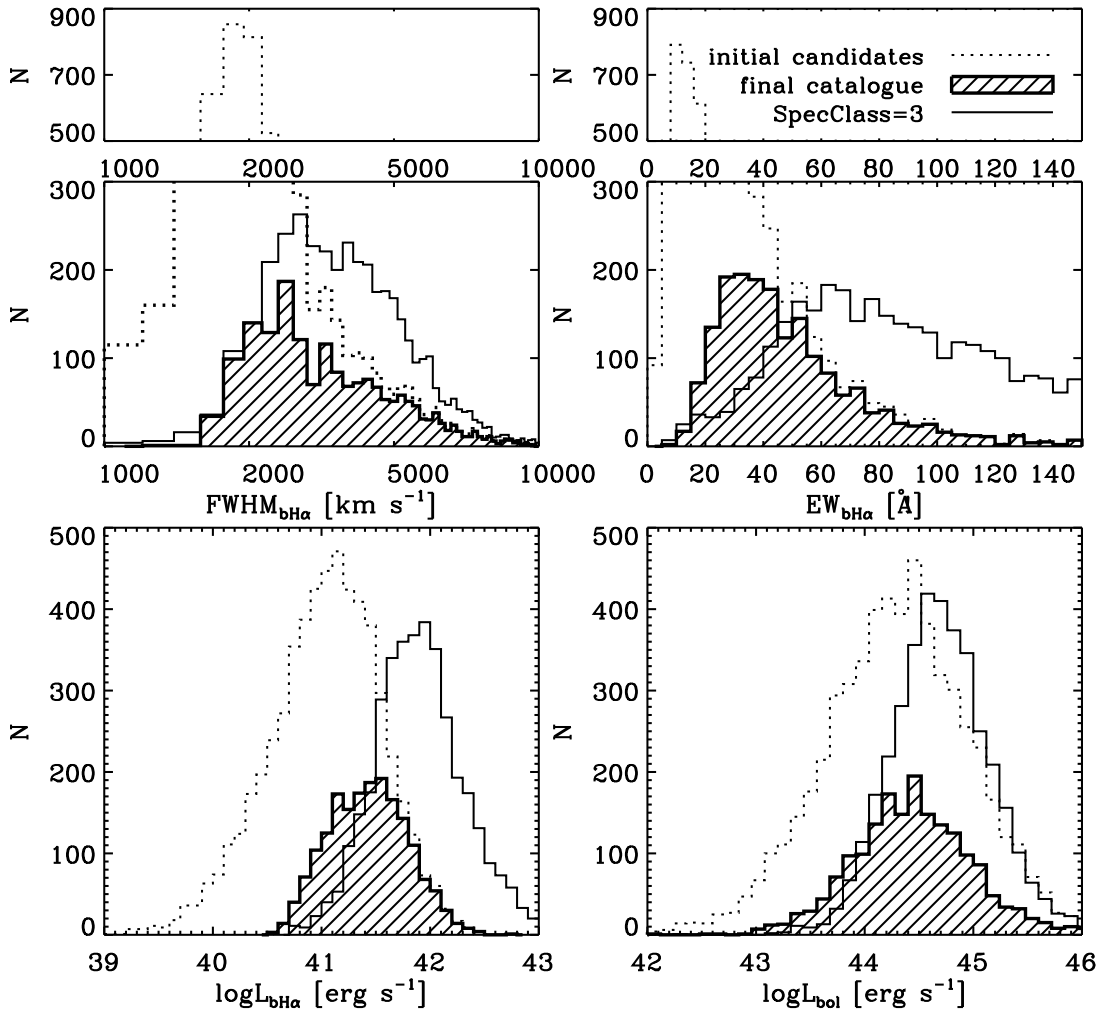} 
\caption{Histograms of FWHM$_{\textrm{bH$\alpha$}}$, EW$_{\textrm{bH$\alpha$}}$, log$L_{\textrm{bH$\alpha$}}$, and log$L_{\textrm{bol}}$ for newly found type 1 AGNs (hatched area, see Section 2.3), \textit{SpecClass} 3 sources (solid lines), and all type 1 AGN candidates (dotted) that were chosen by measuring the flux ratio $\rm{F_{6,533}/F_{6,470}}$ (see Section 2.1).
}
\label{histograms}
\end{figure*}

\begin{figure*}
\centering
\includegraphics[width=0.8\textwidth]{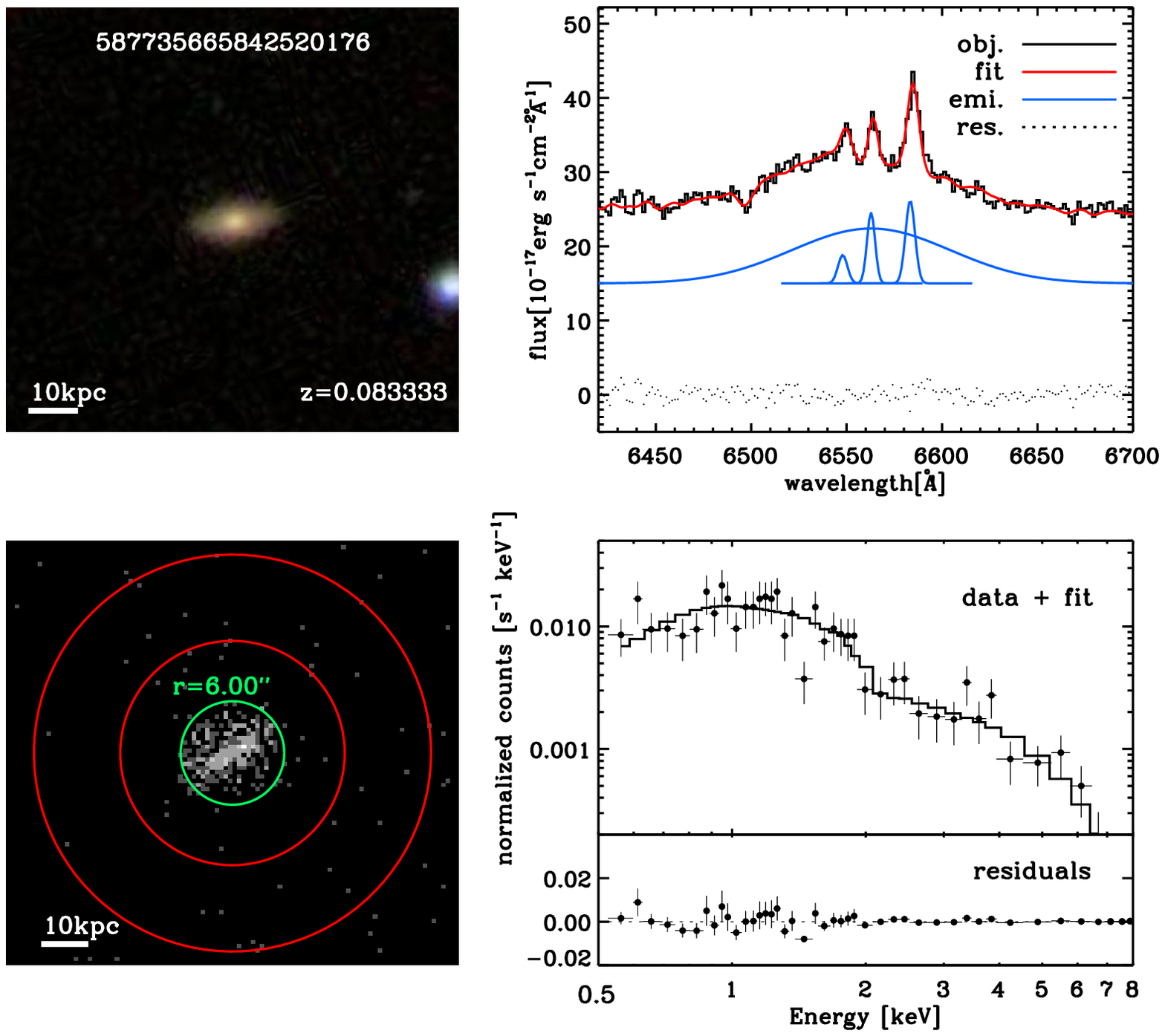} 
\caption{Example \textit{Chandra} X-ray data (obsid = 2779) and model fit, with the SDSS \textit{gri} composite image and our spectral fit. Top left: SDSS \textit{gri} composite image. 
Top right: observed spectrum (black), our fit (red), detected emission lines (blue) including broad components, and residuals (black dots). 
Note that the detected emission lines are arbitrarily shifted towards the y-axis for clarity. 
Bottom left: \textit{Chandra} X-ray image, including source (green circle) and background (red annulus) markers. 
Bottom right: extracted data with the applied fit versus energy (top); residuals (bottom). 
}
\label{SDSS_CHANDRA}
\end{figure*}

\subsection{SDSS \textit{SpecClass}=3 objects}
\label{ssec:specclass}

SDSS DR7 provides a spectroscopic classification of galaxies: \textit{SpecClass}. Broad-line AGNs, especially for low redshift, are flagged \textit{SpecClass}=3 when the observed spectrum satisfies the following criteria: the FWHM of {\it any} emission line is greater than 1,000 \kms, the height of the emission line is at least 3 times higher than its noise (equivalent to our condition of $\rm A/N>3$), and the equivalent width of the line is larger than 10 \AA. In total, there are 4,125 \textit{SpecClass} 3 sources at $z < 0.2$. Galaxies not showing broad-emission lines are classified as \textit{SpecClass} 2. 

We tried applying our type 1 AGN selection technique on the \textit{SpecClass} 3 objects; Tab.~\ref{tab:specclass} summarizes our results. To begin with, 4,079 out of 4,125 objects passed our flux ratio test. Then, 3,969 of the remaining objects passed the A/N and FWHM cuts, and 3,718 (90.13\%) passed the final areal excess test. Overall, our technique rejected roughly 10\% (407 out of 4,125) of the objects.

We describe further why some of the SDSS \textit{SpecClass} 3 objects are not classified the same way in our analysis based on a subsample of the objects that were excluded by the first and most important step of our scheme, the flux ratio test. By performing additional spectral fitting to the 46 ($4,125-4,079$) sources that failed to pass the 1$\sigma$ cut from the flux ratio distribution, we found that 11 objects have empty spectral bins around \Ha\ (see Fig.~\ref{example_spec3_sed}, second row). Hence, our technique could not be feasibly applied to these objects. In the majority of the rest 35 missed objects (25 out of 35), we could not detect any notable broad features; the top row of Fig.~\ref{example_spec3_sed} shows such an example. Of the 46 missed objects, 10 passed our A/N and FWHM tests of \Ha\ and thus possibly have broad \Ha\ components. However, 2 of the 10 failed to pass our 90\% completeness test and a further 3 failed to pass the areal excess cut test; overall, only 5 out of 10 passed all of our test. Similarly, our technique could not be effectively applied on 3 objects because their spectrum around \Ha\ appeared asymmetric and uneven (see Fig.~\ref{example_spec3_sed}, third row). Finally, one object had a spectrum that did not appear to be an ordinary galaxy spectrum, making it difficult to fit this spectrum with a combination of template stellar (population) spectra (Fig.~\ref{example_spec3_sed}, bottom row). Tab.~\ref{tab:specclass} summarizes the statistics of the \textit{SpecClass} 3 sources.

Fig.~\ref{histograms} presents the distributions of the widths and luminosities of the broad \Ha\ component for the newly found type 1 AGNs (hatched area) and \textit{SpecClass} 3 sources (solid line). The new sources show broad \Ha\ widths (FWHM$_{\textrm{bH$\alpha$}}$) comparable to those of the \textit{SpecClass} 3 sources. On the other hand, the characteristic luminosity (i.e., the peak of the distribution) and the equivalent width of their broad \Ha\ components seems lower, by a factor of a few, than that of the \textit{SpecClass} 3 objects, where bolometric luminosity is derived from $L_{\textrm{bol}} \approx$ 3,500\LOIII\ \citep{hec04}. Hence, the newly found objects are mostly {\it low-luminosity} type 1 AGNs. 

\begin{center}
\begin{deluxetable}{ccc}
\tabletypesize{\scriptsize}
\tablecaption{X-ray fitting strategies}
\tablewidth{0pt}
\tablehead{
\colhead{\multirow{2}{*}{count}} &
\multicolumn{2}{c}{fitting prescription} \\
\colhead{} &
\colhead{$N_{\rm H}$ [cm$^{-2}$]\tablenotemark{a}} &
\colhead{$\Gamma$\tablenotemark{b}} 
}
\startdata
$\leq$ 50					& 10$^{20}$ fixed\tablenotemark{c}		& 1.9 fixed	\\
50$< \rm{count} <$100		& 10$^{20-24}$ variable					& 1.9 fixed	\\
$\geq$ 100					& 10$^{20-24}$ variable					& 1.5-2.5 variable	
\enddata
\label{tab:xray_fitting}
\tablenotetext{a}{Hydrogen column density}
\tablenotetext{b}{X-ray photon index}
\tablenotetext{c}{The value is fixed during spectral fit}
\end{deluxetable}
\end{center}

\subsection{\textit{Chandra} X-ray data}
\label{ssec:chandra}
A strong X-ray flux is one of the prominent and characteristic features of an AGN, because an accretion disk surrounding a central black hole produces X-ray emissions \citep{fab89}. Relativistic electrons from the corona in the vicinity of the accretion disk scatter the lower-energy photons radiated from the accretion disk. As a result, the lower-energy photons receive more energy from the electrons (inverse Compton scattering), and this process produces an X-ray spectrum with a power-law continuum \citep{haa91, zdz00, kaw01}. Thus, strong X-ray emission is considered to be evidence of the presence of an AGN. 

We utilized the \textit{Chandra} Source Catalog \citep{eva10} and archival \textit{Chandra} X-ray data to identify X-ray sources among our AGN candidates. We downloaded the matched X-ray images from the \textit{Chandra} archive and applied a 10 arcmin matching radius. Through visual inspection of X-ray images of their optical counterpart, we found 84 X-ray sources among the type 1 AGN samples. The X-ray recovery rate (1.5\%) was very low mainly because X-ray sky coverage was much narrower (by a factor of 25) and depth was much shallower compared to the SDSS optical survey.

\begin{deluxetable*}{ccccccccc}
\tabletypesize{\scriptsize}
\tablecaption{X-ray properties of type 1 AGNs}
\tablewidth{0pt}
\tablecolumns{10}
\tablehead{
\colhead{\multirow{2}{*}{SDSS ObjID}} &
\colhead{\multirow{2}{*}{obsid\tablenotemark{a}}} &
\colhead{RA\tablenotemark{b}} &
\colhead{DEC\tablenotemark{b}} &
\colhead{\multirow{2}{*}{z}} &
\colhead{\multirow{2}{*}{counts}} &
\colhead{count rate} &
\colhead{log$L_{\textrm{2-10keV}}$\tablenotemark{c}} & 
\colhead{\multirow{2}{*}{note\tablenotemark{d}}} \\
\colhead{} &
\colhead{} &
\colhead{[deg]} &
\colhead{[deg]} &
\colhead{} &
\colhead{} &
\colhead{[count ${\rm s}^{-1}$]} &
\colhead{[\ergs]} &
\colhead{}
}
\startdata
587738372207476987 &   3784 &  121.94200 &   21.23800 &         0.14215 &    230 &           0.012 &                  $42.76^{+0.08}_{-0.06}$ &     3    \tabularnewline
587735665842520176 &   2779 &  211.12100 &   54.39800 &         0.08333 &    318 &           0.022 &                  $42.56^{+0.06}_{-0.06}$ &     N    \tabularnewline
587725551741370430 &   2033 &  143.96500 &   61.35300 &         0.03933 &    332 &           0.007 &                  $41.36^{+0.04}_{-0.04}$ &     N    \tabularnewline
587741602568798223 &   4695 &  184.53300 &   28.17600 &         0.17773 &    449 &           0.045 &                  $43.30^{+0.05}_{-0.05}$ &     N    \tabularnewline
587725551197683729 &   7146 &  124.83400 &   50.00700 &         0.13969 &    171 &           0.022 &                  $42.77^{+0.09}_{-0.09}$ &     3    \tabularnewline
587729774220738649 &   9557 &  210.21899 &   -1.75300 &         0.14878 &   1662 &           0.034 &                  $42.92^{+0.02}_{-0.02}$ &     N    \tabularnewline
587731521745518723 &    827 &  140.28600 &   45.64900 &         0.17448 &   4617 &           0.246 &                  $44.16^{+0.02}_{-0.01}$ &    3p    \tabularnewline
587729653959033079 &    887 &  249.20200 &   41.02900 &         0.04737 &    287 &           0.004 &                  $41.28^{+0.05}_{-0.07}$ &     N    \tabularnewline
587735696453009410 &   1623 &  220.76199 &   52.02700 &         0.14121 &   2851 &           0.190 &                  $43.86^{+0.02}_{-0.02}$ &    3p \enddata
\label{tab:chandra}
\tablecomments{(This table is available in its entirety in a machine-readable form in the online journal. A portion is shown here for guidance regarding its form and content.)}
\tablenotetext{a}{\textit{Chandra} observation ID}
\tablenotetext{b}{\textit{Chandra} J2000 coordinates}
\tablenotetext{c}{Errors are estimated with the 90\%\ confidence region for a single interesting parameter.}
\tablenotetext{d}{Discriminating flag for the newly identified type 1 AGNs (`N') and the previously confirmed sources (`3'). Notation 'p' indicates the photon pileup fraction greater than 10\%.}

\end{deluxetable*}

We used the \textit{Chandra} Interactive Analysis of Observations (CIAO) v4.6\footnote{http://cxc.harvard.edu/ciao/} and XSPEC v12.8.1g software\footnote{http://heasarc.nasa.gov/xanadu/xspec/} \citep{arn96} to reprocess the \textit{Chandra} X-ray data and perform a spectral fitting analysis. We manually aligned the coordinates to the actual image center and determined the source aperture, $A_{\rm{s}}$, (green circle in Fig.~\ref{SDSS_CHANDRA}) by reading off the value from the pipeline relation between the off-axis angle and the aperture size suggested. The applied aperture ($\overline{A_{\rm{s}}}=11.2 {\rm{kpc}}$, $\sigma_{A_{\rm{s}}}=10.0 {\rm{kpc}}$) generally covers the whole galaxy ($\overline{\rm{PetroR90_r}}=12.2 {\rm{kpc}}$, $\sigma_{\rm{PetroR90_r}}=6.2 {\rm{kpc}}$, where $\rm{PetroR90_r}$ is the radius containing 90\%\ of Petrosian flux in \textit{r}-band) while some nearby sources represent central part. Then, we used \texttt{specextract} with \texttt{psfcorr} in CIAO which incorporates an energy-dependent aperture correction (average enclosed energy fraction is 85\%). We set the sky background to an annulus of a 10 arcsec width (or 13 arcsec) away from the center when $A_{\rm{s}} < 5$ arcsec (or $A_{\rm{s}} \geq 5$ arcsec). We grouped all spectra with a minimum of 1 count per bin and applied the Cash statistic \citep{cas79} with the assumption of the Poisson distribution to achieve spectral fitting for low-count spectra. XSPEC command \texttt{statistic cstat} applies the W statistic \citep{wac79} for the unmodeled background spectra. We performed spectral fitting by applying the photoelectric absorption model and power law (\texttt{phabs*pow} in XSPEC) in the rest-frame energy range between 0.5 and 8.0 keV. Tab.~\ref{tab:xray_fitting} gives a detailed prescription for the spectral fitting procedure, Tab.~\ref{tab:chandra} describes the X-ray properties measured on these sources, and Fig.~\ref{SDSS_CHANDRA} shows an example result.

\begin{figure}
\centering
\includegraphics[width=0.5\textwidth]{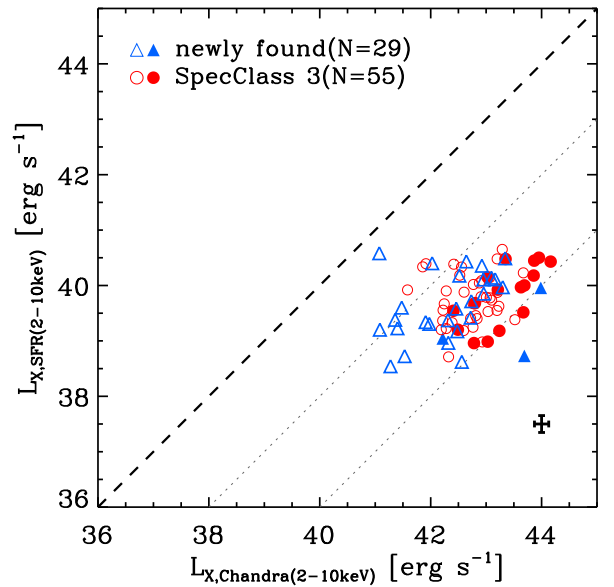} 
\caption{Comparison of 2$\--$10 keV X-ray luminosity between measurements ($L_{\textrm{X, Chandra}}$) and SFR-inferred estimates ($L_{\textrm{X, SFR}}$). Newly found type 1 AGNs and \textit{SpecClass} 3 sources are shown as blue triangles and red circles, respectively. 
 The objects having more than 10\%\ of photon pileup fraction are shown with filled blue triangles (3 out of 29) and filled red circles (16 out of 55).
The black cross at the bottom right indicates the mean error for all sources. 
The central dashed line is a one-to-one fiducial line. Two additional dotted lines are shown to facilitate comparison.
}
\label{Ranalli}
\end{figure}

Apart from AGNs, there are other luminous X-ray sources such as high-mass X-ray binaries, young supernova remnants, and hot ionized interstellar medium resulting from star formation activity. In-depth studies have proposed a correlation between the global star formation rate (SFR) and X-ray luminosity across the whole galaxy \citep{gri03, ran03, grimes05, leh10, min11}. In particular, \citet{ran03} showed the following linear relationship between 2$\--$10 keV X-ray luminosity and SFR based on 23 \textit{bona fide} star-forming galaxies. 
\begin{equation}
	{\rm SFR} [M_{\odot} {\rm yr^{-1}}] = 2.0 \times 10^{-40} L_{\rm 2\--10 keV} 
\end{equation}
We can also estimate SFR by using \Ha\ narrow emission line strength following \citet{ken98}, and then derive the 2$\--$10 keV X-ray luminosity ($L_{\textrm{X, SFR}}$) by using this equation. Fig.~\ref{Ranalli} compares $L_{\textrm{X, SFR}}$ and $L_{\textrm{X, Chandra}}$ for the 84 type 1 AGNs for both the newly found type 1 AGNs (blue triangles) and the SDSS \textit{SpecClass} 3 sources (red circles). Fluxes (horizontal axis) are calculated from the best-fit model. The \textit{Chandra} X-ray luminosities ($L_{\rm X, Chandra}$) are 2$\--$4 orders of magnitude higher than the SFR-inferred estimates ($L_{\textrm{X, SFR}}$), robustly confirming that our new candidates are indeed AGNs.

Regarding photon pileup fraction of our \textit{Chandra} sources, we performed an investigation based on count rate. We explicitly marked the sources with filled symbols in Fig.~\ref{Ranalli} if the object has count rate per second greater than 0.07 which corresponds to a photon pileup fraction greater than 10\%. Loss of photon count underestimates X-ray luminosity (horizontal axis of Fig.~\ref{Ranalli}). There are 19 sources (out of 84) that show photon pileup fraction greater than 10\%, and their intrinsic X-ray luminosity would be found further away from the fiducial line in Fig.~\ref{Ranalli}. Hence, photon pileup does not change our conclusion that 84 X-ray sources are powered by AGN. 

\begin{figure}
\centering
\includegraphics[width=0.5\textwidth]{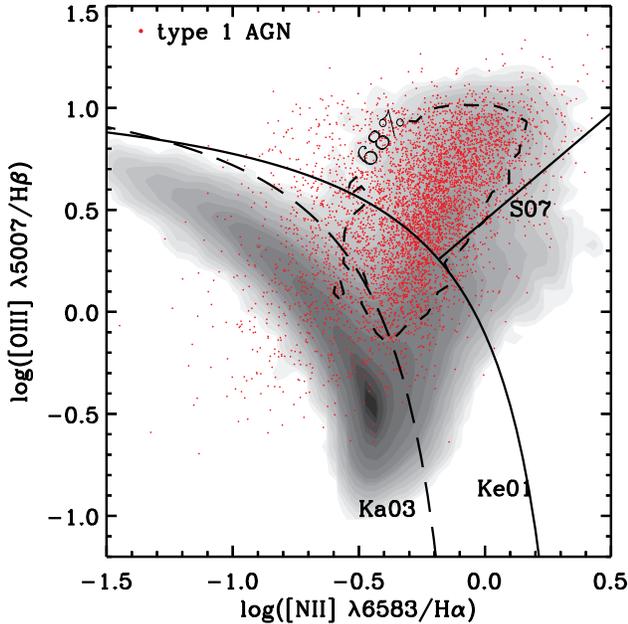} 
\caption{BPT diagnostic diagram for type 1 AGNs (red dots). Emission-line galaxies with $A/N$ cut ($>3$ for \NII, \Ha, \OIII, and \Hb. N$\approx$180,000) chosen from the entire OSSY catalogue are shown with filled contours. Dashed contour indicates 68\%\ distribution for type 1 AGNs.
}
\label{BPT}
\end{figure}

\begin{figure}
\centering
\includegraphics[width=0.5\textwidth]{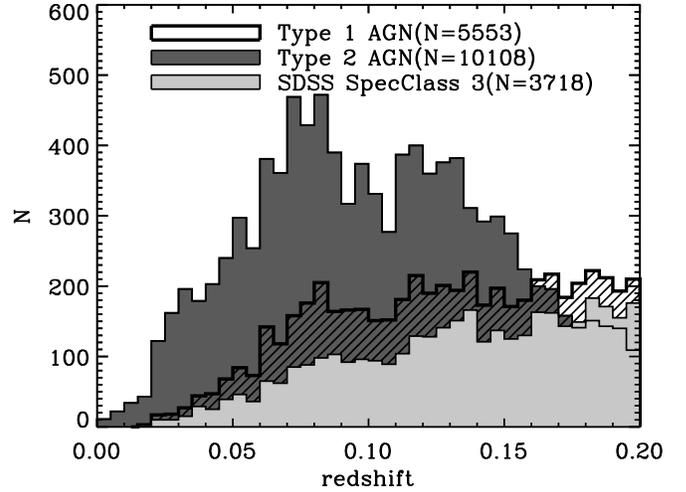} 
\caption{Histograms of type 1 (thick line) and type 2 AGNs (dark gray area) with respect to redshift; the type 1 AGN samples include 3,718 \textit{SpecClass} 3 sources (light gray area). The hatched are represents the newly found type 1 AGNs. 
}
\label{redshift_dist}
\end{figure}

\subsection{Type 2 AGN selection}
Based on the OSSY catalogue (N = 664,187, $z<0.2$), we chose the type 2 AGNs from the Seyfert region of the BPT AGN diagnostic diagram \citep{bal81}. We used the theoretical maximum starburst model of \citet{kew01} and the empirical star formation curve derived by \citet{kau03} as demarcation lines. For the discrimination of low-ionization nuclear emission-line regions from Seyfert AGNs, we used the empirical demarcation of \citet{sch07}. In addition, we used the criterion $A/N>3$ as a threshold to secure statistical significance for the narrow emission lines (\NII, narrow \Ha, \OIII, and \Hb) used in the diagnostics (Fig.~\ref{BPT}). As a result, we selected 10,108 type 2 AGNs at $z<0.2$. The BPT AGN diagnostics diagram for type 1 AGNs is presented in Fig.~\ref{BPT}. Overall distributions of the type 1 and type 2 AGNs versus redshift are shown in Fig.~\ref{redshift_dist}.

\section{Type 1 AGN fraction}
\label{sec:typefraction}

\begin{figure}
\centering
\includegraphics[width=0.5\textwidth]{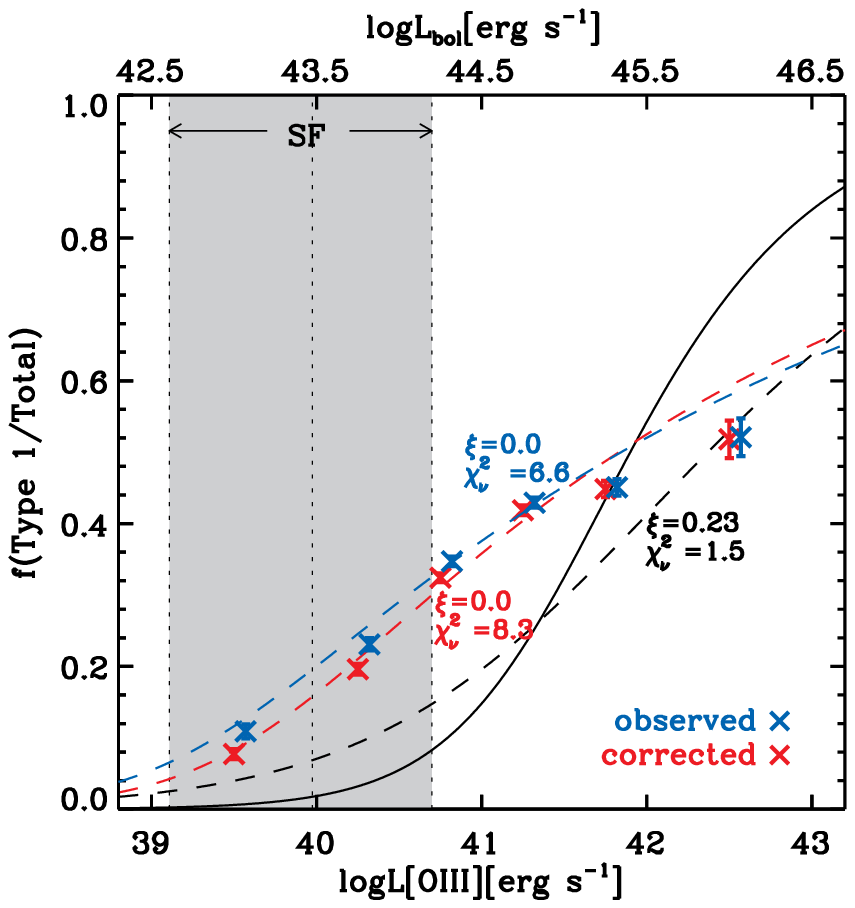} 
\caption{Type 1 AGN fraction with respect to \OIII\ luminosity. 
Bolometric luminosity is shown on the top axis, following $L_{\textrm{bol}} \approx$ 3500\LOIII\ \citep{hec04}.
Blue crosses and error bars show the observed type 1 AGN fraction. 
Red crosses and error bars show the type 1 AGN fraction estimated by statistically correcting for the dilution of observed type 2 AGNs due to star formation.  
The gray band shows the range of \LOIII\ of star-forming galaxies sampled so that their stellar mass distribution matched that of type 2 AGNs. 
Vertical dotted lines denote the median and 95\% distributions.
The black solid curve represents the standard receding torus model proposed by \citet{sim05}. 
The black dashed curve indicates the modified receding torus model which allows to vary the height of the torus according to the nucleus luminosity with $\xi$=0.23. 
The best fits for the observed (blue) and dilution-corrected (red) data points are shown as blue and red dashed lines, including $\xi$. 
Note that the error bars are barely visible due to their small magnitudes, except for the highest \LOIII\ bin. 
}
\label{fraction}
\end{figure}

By combining both types of AGN, we obtained the type 1 AGN fraction, namely the number ratio between type 1 AGNs and all AGNs, as a function of \OIII\ luminosity and bolometric luminosity (Fig.~\ref{fraction}, Tab.~\ref{tab:frac_data}). The type 1 AGN fraction observed (blue crosses in Fig.~\ref{fraction}) increases with \LOIII. This trend is generally consistent with the prediction based on the receding torus model, but does not match it closely: the luminosity dependency of the type 1 AGN fraction is steeper at low \LOIII\ and becomes flattened at high \LOIII\ bins compared to the ``standard'' model (black solid curve) that was derived from the fit to the type 1 AGN fraction from the SDSS DR4. \citet{sim05} noted that the degree of match (to the old data) improved when the height of the torus was modified to depend on the nucleus luminosity in the form of $h \propto L^{\xi}$ (black dashed line) with $\xi > 0$. In the ``modified'' scheme, the sublimation radius of a dust torus gradually increases with the nuclear luminosity \LOIII. In this framework, the type 1 fraction is described as $f_{1} \equiv N_{\rm 1}/N_{\rm tot} = 1-[1+3(L/L_{o})^{1-2\xi}]^{-\alpha}$, where $\alpha$ is set to be 0.5. 
In Simpson's formulation, the best fit is to reproduce $f_{\rm 1} = \alpha$ for a combination of ($L_{\rm 0}$, $\xi$); this yielded $L_{\rm 0} = 42.37\, \rm erg\ s^{-1}$ and $\xi=0.23$ for the reduced $\chi^2_{\nu}$ of 1.5. The modified model does not seem consistent with our new data, however. We will revisit this issue in the next section.

Challenges remain in accurately defining the true type 1 fraction. For example, as discussed in previous sections, our type 1 selection scheme probably excludes some true type 1 AGNs as a consequence of finding robust type 1 AGNs. In this sense, our type 1 fraction may be considered a lower limit. On the other hand, the number of type 2 AGNs is also uncertain. Most notably, we have selected type 2 AGNs only from the Seyfert category based on the BPT diagnostics (see Section 2.5). But, it is likely that there are some type 2 AGNs hidden in the composite and even the star-forming categories. We now attempt to discover them and to include them in our estimation of the type 1 fraction.

\begin{center}
\begin{deluxetable*}{ccccccc}
\tabletypesize{\scriptsize}
\tablecaption{The observed type 1 AGN fraction with the dilution corrected data}
\tablewidth{0pt}
\tablehead{
\colhead{log$L_{\textrm{bol}}$} &
\colhead{log\LOIII} &
\colhead{f(type 1 AGNs/total)} &
\colhead{f(type 1 AGNs/total)\tablenotemark{a}$_{\textrm{corrected}}$} &
\colhead{\multirow{2}{*}{N$_{\textrm{type 1 AGN}}$}} &
\colhead{\multirow{2}{*}{N$_{\textrm{type 2 AGN}}$}} &
\colhead{\multirow{2}{*}{N$_{\textrm{type 2 AGN}}^{\textrm{corrected}\tablenotemark{b}}$}} \\
\colhead{[\ergs]} &
\colhead{[\ergs]} &
\colhead{[\%]} &
\colhead{[\%]} &
\colhead{} &
\colhead{} &
\colhead{}
}
\startdata
$42.5-43.5$	& $39.0-40.0$ & $10.90\pm0.98$ & $7.72\pm0.71$   & 110 & 899   & 416\\ 
$43.5-44.0$	& $40.0-40.5$ & $23.08\pm0.86$ & $19.60\pm0.75$ & 553 & 1843 & 426\\ 
$44.0-44.5$	& $40.5-41.0$ & $34.71\pm0.67$ & $32.42\pm0.63$ & 1778 & 3344 & 363\\ 
$44.5-45.0$	& $41.0-41.5$ & $42.92\pm0.71$ & $41.86\pm0.70$ & 2084 & 2772 & 123\\ 
$45.0-45.5$	& $41.5-42.0$ & $45.08\pm1.16$ & $44.83\pm1.15$ & 833 & 1015 & 10\\ 
$45.5-46.5$	& $42.0-43.0$ & $52.09\pm2.64$ & $51.80\pm2.63$ & 187 & 172 & 2
\enddata
\tablenotetext{a}{Dilution corrected type 1 AGN fraction which is described in the Section 3.1}
\tablenotetext{b}{Number of dilution corrected type 2 AGNs.}
\label{tab:frac_data}
\end{deluxetable*}
\end{center}

\begin{figure}
\centering
\includegraphics[width=0.5\textwidth]{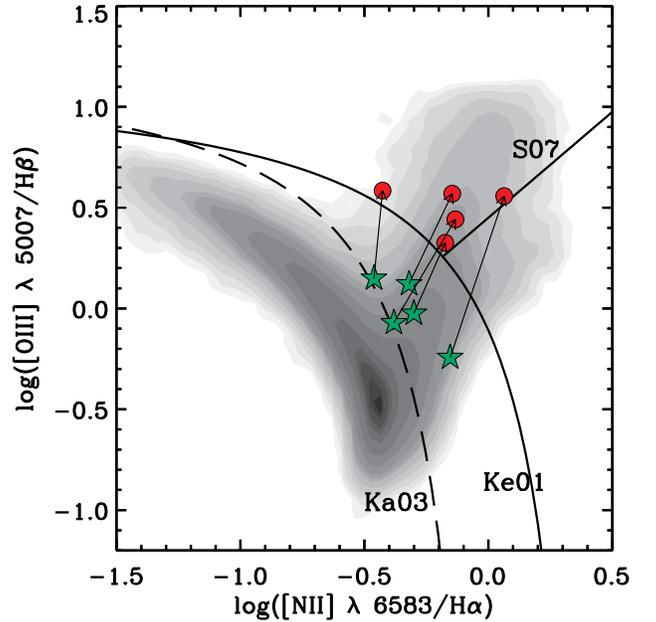} 
\caption{BPT diagnostic diagram with 5 randomly chosen examples. 
All emission line galaxies at $z<0.2$ having $\rm{A/N}>3$ on \NII, \Ha, \OIII, and \Hb\ are represented by filled contours. 
Composite objects (green stars) are shown that were moved to the Seyfert region (red circles) by the procedure described in the main text.
}
\label{BPT_10examples}
\end{figure}

\subsection{Dilution correction on optical lines of type 2 AGNs}
\label{ssec:sfcorrection}

Our analysis starts by setting up two groups of galaxies at $z<0.2$: ``emission-line'' galaxies ($N \sim 200,000$) and star-forming galaxies ($N \sim 150,000$) from the OSSY catalog. Emission-line galaxies obviously include star-forming galaxies. We use only the galaxies with A/N $>$ 3 for all four lines of \NII, \Ha, \OIII\ and \Hb\ that are used in the BPT diagnostics. Assuming that the amount of emission fluxes that come from star formation depends mainly on its stellar mass, we matched every emission-line galaxy with a BPT-selected star-forming galaxy with the same stellar mass within a 5\% tolerance level, based on the stellar mass listed in the MPA-JHU catalog. The size (number of galaxies) of the pool of galaxies with (roughly) the same stellar mass for each galaxy used in our test, which depends strongly on stellar mass itself and the shape of stellar mass function, is roughly 3,400 (median).

This match between an emission-line galaxy and a star-forming galaxy solely based on stellar mass is invalid when it is performed on a single galaxy, because even for the same stellar mass of galaxies, star formation rates can vary widely, resulting in different emission line strengths. However, when this exercise is applied to numerous galaxies, and as long as there is no serious bias of star formation rate in any sense in the sample, the method can be considered largely valid. We subtracted the emission line strength of the star-forming galaxy from the emission-line galaxies, for all four narrow lines, and we adjusted the position of each galaxy in the BPT diagram based on its residual line strengths. Note that we have done this on all emission-line galaxies including star-forming galaxies themselves.

After removing star-forming components in the emission lines, 550 star-forming galaxies were moved to the Seyfert region of the BPT diagram, as shown in Fig.~\ref{BPT_10examples}. This does not mean that there is a significant AGN component in all of these star-forming galaxies. Instead, this arose from the randomness of the choice of our star-forming galaxy as a subtraction. Likewise, after this subtraction, many star-forming galaxies ended up having negative emission lines, again as a result of the randomness of the process. Hence, we decided to ignore this result.

The light emitted by ``composite'' galaxies is believed to include the contributions of both star-forming and AGN activities, and thus this exercise makes more sense when applied to these galaxies; applying this process moved 1,374 composite galaxies from the composite region to the Seyfert region (Fig.~\ref{BPT_10examples}). Again, some of these moves may be a result of applying an unfair star-formation correction due to the randomness of our process. This yielded a total of 11,482 ($10,108+1,374$) type 2 AGNs. The new type 1 AGN fraction that resulted from this ``dilution'' correction is shown in Fig.~\ref{fraction}, using red crosses and error bars. 

The analysis is not complete, because we randomly chose and considered emission line fluxes of star-forming galaxies that only matched the stellar mass. Nevertheless, we can expect a general trend of star-formation-caused dilution of AGN relying on the large-number statistics. The finding of new type 2 AGNs reduces the type 1 AGN fraction, especially in low-luminosity bins, but it seems clear that the current versions of the receding torus model still do not reproduce the ``corrected'' type 1 fraction. 

We now attempt to find fits to our new data by adopting the general form of Simpson's formula, but treating $\alpha$, $L_{\rm 0}$ and $\xi$ as free parameters. We assume $\xi \geq 0$ because this condition is essential to the receding torus model. The best fits were found with $\alpha =0.12$, $L_{\rm 0} = 39.7$, $\xi = 0.0$, and $\chi^2_{\nu}=6.6$ for the ``observed'' data set and with $\alpha=0.14$, $L_{\rm 0} = 40.1$, $\xi = 0.0$, and $\chi^2_{\nu}=8.3$ for the ``corrected'' data set. The new fits are shown as blue and red dashed lines in Fig~\ref{fraction}. However, these two best fits have goodness of fit values ($\chi^2_{\nu}$) that are too large to accept, and are much larger than that found by Simpson for the old data (1.5). In our test, $\chi^2_{\nu}$ became smaller with decreasing $\xi$. Our best fit was found with $\xi =0$. This implies that the introduction of ``modification'' (i.e., a positive luminosity dependence of torus height) seems unsupported. In fact better fits were achieved with a negative value of $\xi$; but the improvement was marginal. The new data call for a new and improved explanation.
\begin{figure*}
\centering
\includegraphics[width=0.9\textwidth]{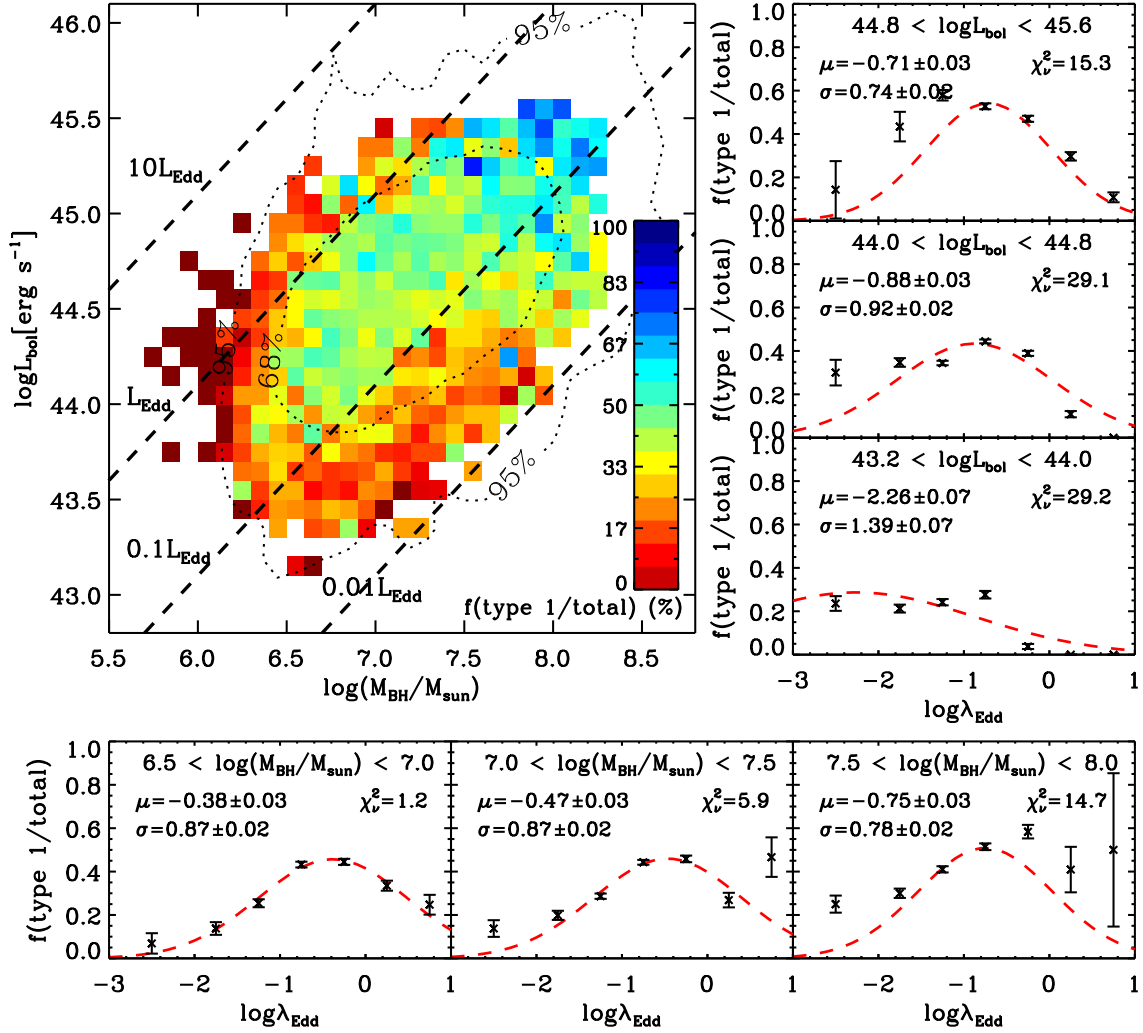} 
\caption{Type 1 AGN fraction with respect to bolometric luminosity and black hole mass. 
The main panel presents the type 1 AGN fraction by using a color grid to denote bins including more than 10 objects in total (type 1 AGNs + type 2 AGNs). Black dotted curves indicate contours (68\%\ and 95\%) for type 1 AGNs, and dashed lines indicate various Eddington ratios. (Right panels) Type 1 AGN fraction versus Eddington ratio for various ranges of bolometric luminosity. 
(Bottom panels) Type 1 AGN fraction versus Eddington ratio for various ranges of black hole mass. 
Red curves in the right and bottom panels show log-normal fits. Tab.~\ref{tab:Lbol_Mbh_table} lists all data points, errors, and fitting results. 
}
\label{Lbol_Mbh}
\end{figure*}

\begin{deluxetable*}{cccccccc} 
\tablecolumns{8} 
\tabletypesize{\scriptsize}
\tablewidth{0pt} 
\tablecaption{Type 1 AGN fraction as a function of Eddington ratio} 
\tablehead{ 
\colhead{}    &  
\multicolumn{3}{c}{log($M_{\rm BH}$/$M_{\odot}$)} 
&   
\colhead{}   & 
\multicolumn{3}{c}{log$L_{\rm bol}$\tablenotemark{a}} \\ 
\cline{2-4} \cline{6-8} \\ 
\colhead{log$\lambda_{\textrm{Edd}}$} & \colhead{$6.5-7.0$}   & \colhead{$7.0-7.5$}    & \colhead{$7.5-8.0$} & 
\colhead{}    & \colhead{$43.2-44.0$}   & \colhead{$44.0-44.8$}    & \colhead{$44.8-45.6$}
}
\startdata 
$-3.0- -2.0$ 	& $0.07\pm0.05$	&	$0.14\pm0.04$	&		$0.25\pm0.04$		& & $0.24\pm0.03$	&	$0.30\pm0.06$	& $0.14\pm0.13$ \\
$-2.0- -1.5$ 	& $0.14\pm0.03$	&	$0.20\pm0.02$	&		$0.30\pm0.02$ 	& & $0.21\pm0.02$	&	$0.35\pm0.02$	& $0.43\pm0.07$ \\
$-1.5- -1.0$ 	& $0.25\pm0.02$	&	$0.29\pm0.01$	&		$0.41\pm0.01$ 	& & $0.24\pm0.01$	&	$0.34\pm0.01$	& $0.58\pm0.02$ \\
$-1.0- -0.5$ 	& $0.43\pm0.01$	&	$0.44\pm0.01$	&		$0.51\pm0.02$ 	& & $0.28\pm0.02$	&	$0.44\pm0.01$	& $0.53\pm0.01$ \\
$-0.5- 0.0$ 	& $0.44\pm0.01$	&	$0.46\pm0.02$	&		$0.58\pm0.03$ 	& & $0.04\pm0.01$	&	$0.39\pm0.01$	& $0.47\pm0.01$ \\
$0.0- 0.5$  	& $0.33\pm0.02$	&	$0.27\pm0.03$	&		$0.41\pm0.10$ 	& & $0.00\pm0.00$	&	$0.11\pm0.01$	& $0.30\pm0.02$ \\
$0.5- 1.0$  	& $0.25\pm0.05$	&	$0.47\pm0.09$	&		$0.50\pm0.35$ 	& & $0.00\pm0.00$	&	$0.00\pm0.00$	& $0.11\pm0.02$ \\
 \cutinhead{fit with log-normal distribution}
 $\mu$				& $-0.38\pm0.03$	&	$-0.47\pm0.03$	&	$-0.75\pm0.03$	& &	$-2.26\pm0.07$	&	$-0.88\pm0.03$	&	$-0.71\pm0.03$ \\
 $\sigma$				& $0.87\pm0.02$	&	$0.87\pm0.02$		&	$0.78\pm0.02$		& &	$1.39\pm0.07$		&	$0.92\pm0.02$		&	$0.74\pm0.02$ \\
 $\chi^{2}_{\nu}$	& 1.2			&	5.9				&	14.7				& &	29.2				&	29.1				&	15.3
\enddata
\tablenotetext{a}{in unit of [\ergs]}
\label{tab:Lbol_Mbh_table}
\end{deluxetable*}

\begin{figure}
\centering
\includegraphics[width=0.48\textwidth]{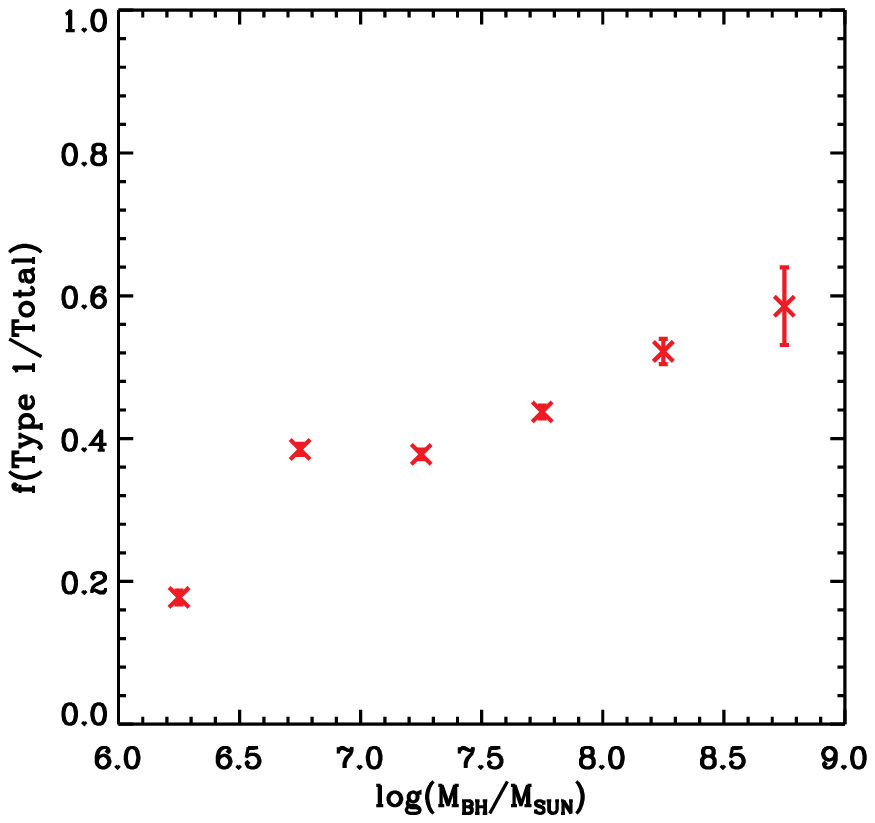} 
\caption{Type 1 AGN fraction with respect to black hole mass.
}
\label{Mbh_fraction}
\end{figure}

\section{Summary \& Discussion}
\label{sec:discussion}

We have presented a process for finding new type 1 AGNs at $z<0.2$ from the SDSS DR7 database. Adding the new type 1 AGNs (1,835) to those in the previous database (3,718) that also pass our selection criteria increases the size of the type 1 AGN database by 49\%. Based on the new database, we calculated a new type 1 AGN fraction, which provides a useful test and constraint to the theoretical understanding of the AGN geometry. We are pleased to release the spectroscopic properties of all type 1 AGNs, given in Tab.~\ref{tab:all_type1}, and our calculations of the type 1 AGN fraction based on \LOIII\ and the Eddington ratio, given in Tab.~\ref{tab:frac_data} and Tab.~\ref{tab:Lbol_Mbh_table}, respectively. It is our wish that these tables will be useful to the community. 

We showed that our updated type 1 AGN fraction is not consistent with the current versions of the receding torus model. The type 1 AGN fraction in the low and intermediate luminosity bins is significantly higher than that given by previous data and models, even after correcting for dilution effects on emission lines caused by star formation.  We found fits to the new data based on the basic formulation of the receding torus model, allowing large differences in the parameterization process, but these fits were poor. Thus, the new fits do not seem to support the hypothesis that torus depth depends on AGN luminosity.

We present the type 1 AGN fraction with respect to black hole mass and bolometric luminosity in Fig.~\ref{Lbol_Mbh}. We used the same bolometric correction factor, 3,500, to both types of AGN following \citet{hec04}. The bolometric correction factor may be uncertain; but if there is no serious type bias on it, our results and conclusions in the relative sense would hold good. The color grids in the main panel of the Fig.~\ref{Lbol_Mbh} show that there is a favored area in which the type 1 AGN fraction is high, for higher black hole mass and bolometric luminosity. This ridge-shaped distribution suggests that neither black hole mass nor bolometric luminosity solely determines the type 1 AGN fraction. Rather, the type 1 AGN fraction suggests a rising and falling distribution versus the Eddington ratio (see small panels in Fig.~\ref{Lbol_Mbh}). If this trend is real, it may imply that the opening angle initially widens with increasing Eddington ratio until the Eddington limit, and diminishes thereafter, which is not expected based on the basic scheme of the receding torus model. Admittedly, however, the goodness of fits ($\chi^2_{\nu}$) is too poor to draw any definite conclusion yet. 

The structure of the dust torus is a key ingredient of understanding the AGN unification theory. However, its physical properties and the detailed geometry are still under debate. \citet{kro88} proposed that the dust and gas in the torus take the form of optically thick, self-gravitating clouds. A clumpy and obscuring torus requires a small volume filling factor and a sufficient scale height compared to the mean freepath of clouds \citep{hon07}. In the framework of a clumpy torus model, the largest clouds in the torus become gravitationally unbound from the high-luminosity AGN when the AGN accretes and radiates close to the Eddington limit. Based on this argument, the relationship $H/R\sim L^{-1/4}$ was derived where $H$ is the scale height, $R$ is inner radius of the torus, and $L$ is the AGN luminosity. Note here that the scale height inversely correlates with the AGN luminosity unlike in the receding torus model. \citet{eli06} described the origin of dust clouds as an outflow wind released from an accretion disk. Because outflow velocities are expected to decrease with $R$, the torus outflow scenario also predicts luminosity dependency of the type 1 AGN fraction. Again, however, our data shown in Fig.~\ref{Lbol_Mbh} do not seem to show a monotonic luminosity dependence. Rather, the covering factor of the torus seems to vary with the Eddington ratio, showing a favored spot in the black hole mass and bolometric luminosity diagram. 

From the perspective of black hole mass, i.e., horizontal axis of Fig.~\ref{Lbol_Mbh}, smaller black hole mass regime shows lower type 1 AGN fraction. It can clearly be seen in Fig.~\ref{Mbh_fraction} that type 1 AGN fraction gradually increases with black hole mass. As lower mass black hole allows more orbits close to the centre, the accretion disk is more heated up overionising broad \Ha\ emitting BLRs. However, it should be noted that further detailed analysis is required to confirm this interpretation. 

It is also noteworthy that there are discussions regarding the effect of the Eddington ratio and luminosity of AGN on the presence of BLRs. \citet{nic03} presented evidence that hidden BLRs are found above the certain threshold value of Eddington ratio based on their earlier model \citep{nic00} and a handful of spectropolarimetric samples with X-ray data. They claimed that BLRs are formed by accretion disk instabilities, and the critical radius becomes smaller than the innermost stable orbit under the low enough accretion rates. Similarly, \citet{eli09} showed that type 1 AGNs are not found below a certain luminosity based on disk-wind scenario. Their data supports the disappearance of BLRs at radiatively inefficient accretion regime which is far more lower than our data presented.

In conclusion, our new type 1 AGN catalogue seems to pose a serious challenge to the current favorite versions of the AGN unified model, but the improved database will hopefully help us better understand the AGN classifications in the end.

\begin{deluxetable*}{ccccclcccccl}
\tabletypesize{\scriptsize}
\tablecaption{Properties of type 1 AGNs\tablenotemark{a}}
\tablewidth{0pt}
\tablecolumns{12}
\tablehead{
\colhead{\multirow{2}{*}{SDSS ObjID}} &
\colhead{RA\tablenotemark{b}} &
\colhead{DEC\tablenotemark{b}} &
\colhead{\multirow{2}{*}{z}} &
\colhead{log\LOIII} &
\colhead{FWHM$_{\textrm{bH$\alpha$}}$} &
\colhead{EW$_{\textrm{bH$\alpha$}}$} &
\colhead{log$L_{\textrm{bH$\alpha$}}$} &
\colhead{\multirow{2}{*}{log($M_{\textrm{BH}}$/$M_{\odot})$\tablenotemark{c}}} &
\colhead{log$L_{\textrm{bol}}$\tablenotemark{d}} &
\colhead{\multirow{2}{*}{notes\tablenotemark{e}}} \\
\colhead{} &
\colhead{[deg]} &
\colhead{[deg]} &
\colhead{} &
\colhead{[\ergs]} &
\colhead{[\kms]} &
\colhead{[\AA]} &
\colhead{[\ergs]} &
\colhead{} &
\colhead{[\ergs]} &
\colhead{} 
}
\startdata
587728309632106568 &       152.29813 &         2.74760 &         0.06319 &           39.98 &  $6671.6\pm   89.2$&    56.9 &   41.18 &    7.55 $^{+   0.05}_{-   0.05}$ &           43.52 &    N \\
587729229300760654 &       244.32758 &        52.48399 &         0.06319 &           39.85 &  $3857.1\pm  159.1$&    18.6 &   40.77 &    6.83 $^{+   0.06}_{-   0.05}$ &           43.40 &    N \\
587724648182448168 &       178.02454 &        -3.50442 &         0.06320 &           41.46 &  $4120.9\pm   23.7$&   170.6 &   42.10 &    7.62 $^{+   0.05}_{-   0.05}$ &           45.00 &    3 \\
587732482763915309 &       229.21861 &        39.90373 &         0.06323 &           40.74 &  $4606.0\pm  167.3$&    18.9 &   41.06 &    7.15 $^{+   0.06}_{-   0.06}$ &           44.29 &    N \\
587742061069074446 &       175.31734 &        21.93939 &         0.06323 &           42.26 &  $3850.9\pm   17.2$&   204.5 &   42.38 &    7.72 $^{+   0.06}_{-   0.05}$ &           45.81 &    3 \\
588295842853159031 &       156.47778 &        47.24015 &         0.06325 &           40.50 &  $3736.5\pm   95.4$&    26.7 &   40.96 &    6.91 $^{+   0.05}_{-   0.04}$ &           44.05 &   XN \\
587742188840943769 &       208.48332 &        21.99855 &         0.06333 &           40.39 &  $6186.1\pm  101.2$&    41.7 &   41.34 &    7.57 $^{+   0.06}_{-   0.06}$ &           43.93 &    N \\
587731522275311780 &       123.89431 &        35.28634 &         0.06337 &           40.75 &  $3480.6\pm   39.1$&    64.5 &   41.36 &    7.07 $^{+   0.04}_{-   0.04}$ &           44.29 &    3 \\
587739097528270899 &       190.37257 &        37.36722 &         0.06340 &           41.26 &  $4243.6\pm   32.3$&   118.5 &   41.84 &    7.50 $^{+   0.05}_{-   0.05}$ &           44.81 &    3 \\
588023668097875996 &       183.71655 &        19.01166 &         0.06340 &           40.37 &  $4734.6\pm   75.3$&    57.1 &   41.19 &    7.25 $^{+   0.05}_{-   0.04}$ &           43.91 &    3 \\
587739827668844624 &       224.67809 &        21.60277 &         0.06343 &           38.78 &  $2343.2\pm    9.9$&   243.0 &   42.18 &    7.16 $^{+   0.04}_{-   0.04}$ &           42.32 &    3 \\
587730847966888351 &       326.21548 &         0.47415 &         0.06344 &           40.62 &  $3770.0\pm   83.4$&    34.5 &   40.89 &    6.88 $^{+   0.04}_{-   0.04}$ &           44.17 &    G \\
587742060531089515 &       172.65646 &        21.23771 &         0.06347 &           40.14 &  $4781.4\pm  149.6$&    24.6 &   40.99 &    7.14 $^{+   0.06}_{-   0.05}$ &           43.68 &    G \\
587741720679612540 &       203.14915 &        25.29781 &         0.06353 &           39.93 &  $4987.1\pm   99.6$&    52.1 &   41.23 &    7.32 $^{+   0.05}_{-   0.05}$ &           43.48 &    N \\
587734690883698778 &       127.78179 &         5.35165 &         0.06356 &           40.93 &  $4447.3\pm   65.3$&    55.8 &   41.34 &    7.27 $^{+   0.05}_{-   0.05}$ &           44.48 &   3G \\
587739721376792749 &       225.65990 &        24.22051 &         0.06361 &           41.11 &  $4392.0\pm   43.7$&    92.6 &   41.42 &    7.31 $^{+   0.04}_{-   0.04}$ &           44.66 &   3G \\
587741600421314588 &       184.55672 &        26.40504 &         0.06361 &           40.21 &  $2131.0\pm   20.9$&    70.4 &   41.17 &    6.52 $^{+   0.02}_{-   0.02}$ &           43.75 &   3G \\
588017977836765242 &       232.13504 &        28.96422 &         0.06362 &           40.72 &  $4744.4\pm   64.9$&    61.1 &   41.42 &    7.38 $^{+   0.05}_{-   0.05}$ &           44.27 &    G \\
587739810484650051 &       208.23201 &        25.48323 &         0.06366 &           41.30 &  $2505.1\pm   14.0$&   175.1 &   41.84 &    7.03 $^{+   0.03}_{-   0.03}$ &           44.84 &  3HG 
\enddata
\label{tab:all_type1}
\tablecomments{(This table is available in its entirety in a machine-readable form in the online journal. A portion is shown here for guidance regarding its form and content.)}
\tablenotetext{a}{This table is also available in the OSSY web site (http://www.gem.yonsei.ac.kr/ossy/))}
\tablenotetext{b}{J2000 coordinates}
\tablenotetext{c}{Black hole mass derived from the method developed by \citet{gre05}}
\tablenotetext{d}{Bolometric luminosity derived from the method developed by \citet{hec04}}
\tablenotetext{e}{Flag presenting the newly identified type 1 AGNs (`N') with the \textit{Chandra} X-ray data (`X'). 
`3' indicates the SDSS \textit{SpecClass} 3 objects. `H' and `G' indicate the object also identified by \citet{hao05} and \citet{gre07}, respectively.}
\end{deluxetable*}

\section*{Acknowledgments}

We are grateful to the anonymous referee for a number of clarifications that improved the quality of the manuscript. 
We also thank Kurt Soto for his valuable comments. 
KO acknowledges support from Swiss Government Excellence Scholarship for the academic year $2013\--2014$ (ESKAS No.2013.0308). 
SKY acknowledges support from the National Research Foundation of Korea (Doyak 2014003730). 
SKY acted as the corresponding author. 
KS and MK gratefully acknowledge support from Swiss National Science Foundation Professorship grant PP00P2 138979/1 and MK support from SNSF Ambizione grant PZ00P2 154799/1. 
This study was performed under the DRC collaboration between Yonsei University and the Korea Astronomy and Space Science Institute. 

\bibliographystyle{yahapj}

\clearpage
\end{document}